\documentclass[preprint]{osa-article}
\journal{osajournal}
\articletype{Research Article}

\usepackage{lineno}
\usepackage{savesym}

\savesymbol{tablenum}
\usepackage[output-decimal-marker={.}, exponent-product=\cdot, per-mode=fraction]{siunitx}
\restoresymbol{SIX}{tablenum}

\usepackage{graphicx}
\graphicspath{{figures/}}

\usepackage{bm}

\usepackage{tabularx}
\usepackage{multirow}

\newcommand{\abs}[1]{\left| #1 \right|}
\DeclareSIUnit[]\rsun
{\text{R\ensuremath{_{\textup{sun}}}}}

\begin{document}

%\shorttitle{A methodology to derive the PSF from partially occulted images}
%\shortauthors{Hofmeister et al.}

\title{Deriving instrumental point spread functions from partially occulted images}

\author{Stefan J. Hofmeister\authormark{1,2,*}, Michael Hahn\authormark{1}, Daniel Wolf Savin\authormark{1}}

\address{\authormark{1}Columbia Astrophysics Laboratory, Columbia University, 550 West 120th Street, New York, New York 10027, USA\\
\authormark{2}Leibniz Institute for Astrophysics Potsdam, An der Sternwarte 16, 14482 Potsdam, Germany}

\email{\authormark{*}stefan.hofmeister@web.de} %% email address is required

\begin{abstract}
The point-spread function (PSF) of an imaging system describes the response of the system to a point source. Accurately determining the PSF enables one to correct for the combined effects of focussing and scattering within the imaging system, and thereby enhance the spatial resolution and dynamic contrast of the resulting images. 
We present a semi-empirical semi-blind methodology to derive a PSF from partially occulted images. We partition the two-dimensional PSF into multiple segments, set up a multi-linear system of equations, and directly fit the system of equations to determine the PSF weight in each segment. The algorithm is guaranteed to converge towards the correct instrumental PSF for a large class of  occultations, does not require a predefined functional form of the PSF, and can be applied to a large variety of partially occulted images, such as within laboratory settings, regular calibrations within a production line or in the field, astronomical images of distant clusters of stars, or partial solar eclipse images. We show that the central weight of the PSF, which gives the percentage of photons that are not scattered by the instrument, is accurate to bettern than \SI{1.2}{\percent}. The mean absolute percentage error between the reconstructed and true PSF is usually between \num{0.5}~and \SI{5}{\percent} for the entire PSF, between \num{0.5}~and \SI{5}{\percent} for the PSF core, and between \num{0.5}~and \SI{3}{\percent} for the PSF tail.
\end{abstract}

\section{Introduction}

The resolution and dynamic response of an imaging system, be it for light or for matter waves, is ideally given by the diffraction limit the system. In practice, it is further degraded by any distortions due to the imaging components and by the internal structure of the imaging system, which together we dub imperfections. These imperfections can arise from the frequency dependence of the refractive index of the imaging components, the microroughness of mirrors or lenses, design flaws such as internal reflections or a misalignment of the imaging components, or from additional imaging analysis components such as optical filters. The point-spread function (PSF) of the imaging system, equal to the impulse response of the instrument to a point source, describes the combined effects of the focusing properties of the imaging system along with all the imperfections of the system. Accurately determining the PSF enables one to correct for these imperfections and can significantly improve the effective resolution and dynamic contrast of the imaging system.

Reliably determining the PSF of an imaging system is still a challenging task. There are three main approaches for determining a PSF. First, one can create a computational model of the entire imaging system with all its components. This requires accurate and detailed knowledge of the physical instrument and enables one to build a quantitative model for the PSF. As the PSF depends on the focus selected, this task is usually only performed for static imaging systems such as remote observing instruments on satellites. However, it can account neither for unknown imperfections of the imaging system nor for its temporal evolution due, for instance, to exposure to the radiation field in interplanetary space. Instruments for which this has been performed include the Hubble space telescope \cite{krist1993, krist1995, hasan1995}, Chandra \cite{jerius2000, karovska2001, carter2003},  the Atmospheric Imaging Assembly (AIA) onboard the Solar Dynamics Observatory \cite{grigis2012}, and the Nuclear Spectroscopic Telescope Array \cite{westergaard2012, madsen2015}.

Second, one can empirically determine the PSF by observing the impulse response of a sup-pixel point source. Such point sources can be quantum dots or sub-resolution fluorescent beads in laboratory settings, or distant stars or artificial laser guide stars in astronomy. This empirical PSF represents the PSF at a specific time and can be either directly used to correct images for scientific analysis or to calibrate a theoretical model of the imaging system \cite{hiraoka1990, shaw1991, boutet2001, Juskaitis2006, Jizhou2018, madsen2015, jerius2000}. Determining the PSF empirically provides a good estimate of the PSF core, which describes the short-distance scattered light and is related to image blurring. But the intensity of a point source is usually too low to enable one to measure the tail of the PSF accurately, which correspond to the long-distance scattered fraction of the light. Although long-distance scattering is several orders of magnitudes weaker than short-distance scattering, the accumulated effect over the field of view can result in significant offset intensities in the darkest image regions. Furthermore, scattering over the entire image can also significantly reduce peak intensities of very bright point-like sources in the image. Both effects reduce the dynamic contrast of the image and result in incorrect image intensities. 

Lastly, one can reconstruct the PSF by comparing an observed image with the true image, i.e., an image unaffected by instrumental effects of the imaging system. The true image, however, is in general not known. Therefore, blind reconstruction techniques have been developed to simultaneously reconstruct the true image together with the true PSF from a single observed image. This process is ill-posed and usually does not have a unique solution. The result depends on the first guess of the PSF and the blind reconstruction algorithm chosen. Typically the algorithms constrain the inversion problem by adding a priori information in the form of a regularization parameter, such as Tikhonov's regularization \cite{tikhonov1977} or the total variance regularization \cite{rudin1992, rudin1994}. These algorithms are mostly fast and can significantly improve the image quality. However, since their solution is not unique and the a priori information employed is an assumption, the reconstructed PSF is not necessarily the correct instrumental PSF. One needs to take great care when using such a PSF to correct other images for instrumental effects.

Here, we present a semi-empirical semi-blind deconvolution methodology to accurately determine the instrumental PSF from images where the true intensities in some image pixels are known a priori to be zero\footnote{A Python implementation of this algorithm is available at https://github.com/stefanhofmeister/Deriving-PSFs-from-partially-occulted-images}. This zero-intensity information can come from an occulter in the object focal plane, from external occultations where the occulter and the object are both at large distances from the image plane (e.g., solar eclipses), or from pixels for which the true intensities are known a priori to be negligible without having an external occultation (e.g., pixels in astronomical images which do not contain a star). We dub all these cases as partially occulted images. 
Using partially occulted images has several advantages. Having only a portion of the image occulted, we are able to employ dramatically more photons compared to point-source-based empirical methodologies. This enables our algorithm to fit both the core and the tail of the PSF with a high level of precision. As the true intensities in the occulted region are known a priori to be zero, our algorithm is not entirely blind. This enables our algorithm to converge towards the correct instrumental PSF. Additionally, a priori knowledge on the shape of the PSF, such as from any observed Fraunhofer diffraction patterns from metal-mesh optical filters, can easily be incorporated. Furthermore, our approach enables one to revise existing high-quality PSFs derived from other approaches, to test their fidelity, and to add missing components. Thus, the algorithm is an advancement based on empirical point-source-based methods and blind deconvolution methods. The rest of the paper is structured as follows. In Section~\ref{sec:concept}, we first describe the mathematical concept of the algorithm. In Section~ \ref{sec:implementation}, we give details regarding the implementation. In Section~\ref{sec:evaluation}, we showcase the convergence for several test cases, and in Section~\ref{sec:prospects}, we summarize our results.

\section{Concept} \label{sec:concept}
In the following sections, we describe the mathematical concept of our algorithm, how it can be used to derive or revise a PSF from partially occulted images, and under which conditions our algorithm is guaranteed to converge towards the true PSF.

\subsection{General} \label{sec:general}
The PSF is equivalent to the instrumental impulse response to a point source. With $I_\text{t}$ being the true intensity of a point source at a location $\textbf{r$_\text{t}
$}$ in the image plane, the observed intensity $I_\text{o}$ within an infinitesimally small image segment $dA$ at a location $\pmb{r}$ in the image plane is given by
\begin{align}
       I_\text{o,$\pmb{r}$} &= I_\text{t,$\pmb{r_\text{t}}$}\ \text{psf}(\pmb{r}-\pmb{r_\text{t}})\ dA + \epsilon  \nonumber \\ 
       &=  I_\text{t,$\pmb{r}$-$\pmb{\Delta r}$}\ \text{psf}(\pmb{\Delta r})\ dA+ \epsilon, \label{eq:defpsf}
\end{align}
where $\text{psf}(\pmb{r} - \pmb{r_\text{t}})$ is the instrumental scattering function giving the fractional number of photons that are scattered from their expected location $\pmb{r_\text{t}}$ in the image plane into an area of size $dA$ located at $\pmb{r} - \pmb{r_\text{t}}$ in the image plane, $\pmb{\Delta r} = \pmb{r} - \pmb{r_\text{t}}$, and $\epsilon$ is a noise component in the observed signal.
The integral of the PSF over the entire image plane $P$ is one, 
\begin{equation}
\iint_{P} dA\ \text{psf}(\pmb{\Delta r}) =1, \label{eq:nophotonslost}
\end{equation}
as long as we assume perfect reflectivity of the optical components, i.e., that no photons are lost.

We interpret an image as the superposition of numerous point sources and approximate the PSF to be shift-invariant, i.e., that the PSF does not depend on the location in the image plane. The observed intensity in an image pixel is given by $I_\text{t,\pmb{r}} \ \text{psf}(\pmb{\Delta r}=0)$ plus the scattered light from all other point sources located at distances $\abs{\pmb{\Delta r}}>0 $ from the observed pixel, i.e., 
\begin{align}
I_\text{o,$\pmb{r}$} &= \iint_P dA \ I_\text{t,$\pmb{r}$-$\pmb{\Delta r}$} \ \text{psf}(\pmb{\Delta r}) + \epsilon  \nonumber \\
 &= \sum_{\pmb{\Delta r}} I_\text{t,$\pmb{r}$-$\pmb{\Delta r}$} \ \text{psf}_{\pmb{\Delta r}} + \epsilon  \nonumber \\
  &= \sum_{S}  \text{psf}_S \sum_{\pmb{\Delta r} \text{ in } S}  I_\text{t,$\pmb{r}$-$\pmb{\Delta r}$} + \epsilon . \label{eq:psf}
\end{align}
In the second line of Equation~\ref{eq:psf}, we have discretized the PSF in the summation into segments having the size of one image pixel. The summation goes over the entire image plane described by the vector $\pmb{\Delta r}$. The coefficients $\text{psf}_{\pmb{\Delta r}}$ give the number of photons which are scattered into the direction $\pmb{\Delta r}$ into one pixel. In the third line, we have discretized the PSF into segments $S$ by aggregating PSF coefficients $\text{psf}_{\pmb{\Delta r}}$ over regions where the PSF varies slowly; $\text{psf}_S$ denotes the mean PSF weight of the segment. This segmentation is an approximation and converges to the exact solution when the size of the PSF segments $S$ become sufficiently small. The error involved in this approximation is negligible compared to other errors in our methodology that we describe below. As the PSF decreases rapidly within the core of the PSF and slowly in its tail, one would typically choose small PSF segments in the PSF core region and larger segments in the PSF tail region. 

Equation~\ref{eq:psf} defines the main concept of our algorithm. Under the presumption that both the true image $I_\text{t}$ and the observed image $I_\text{o}$ are known very accurately, that the noise level $\epsilon$ is small, and that there are more observed pixels $I_{\text{o},\pmb{r}}$ than PSF coefficients $\text{psf}_S$, the PSF coefficients $\text{psf}_S$ can be determined by a multi-linear fit between the true and observed intensities. For the reminder of this section, we presume the noise level to be negligible, which is further discussed in Section~\ref{sec:implementation}.

Next, we investigate the requirements on suitable images for determining the PSF in laboratory settings. There are two trivial kinds of images where we know the true intensities in advance: (1)~occulted images, and (2)~images with a uniform intensity. In the first case, the true intensities in the occulted area are zero. In the second case, the true intensity of each pixel is equal, and Equation~\ref{eq:psf} simplifies to 
\begin{equation}
    I_\text{o,$\pmb{r}$} = I_\text{t,$\pmb{r}$} \sum_{\pmb{\Delta r}} \text{psf}_{\pmb{\Delta r}} = I_\text{t,$\pmb{r}$} . \label{eq:IoIt}
\end{equation}
Thus, the observed intensity is the true intensity for uniformly illuminated images. This is because each image location in a uniformly illuminated image has scattered away exactly the same number of photons away as it receives from all other locations. Equation~\ref{eq:IoIt} is also approximately true for partial image illuminations as long as the uniformly illuminated area is much larger than the area of influence of the PSF. Only close to the edge of the illuminated area do the true and observed intensities start significantly to differ. 
Combining these two cases, if follows that the true intensities are also well known for large parts of an image that is partially occulted and that is uniformly illuminated in the unocculted area. 

We construct a first estimate for the true image of such a partially occulted image by setting the intensities in the occulted pixels to zero. The observed intensities in the occulted pixels can now be used to fit for an estimate of the PSF coefficients $\text{psf}_{\pmb{\Delta r}>0}$ using Equation~\ref{eq:psf}, and $\text{psf}_{\pmb{\Delta r}=0}$ can be derived from Equation~\ref{eq:nophotonslost}. The core of the PSF will not be fitted accurately by this first fit, as the true intensities near the illuminated edge are not well known in advance. This issue can be resolved by an iterative approach. Using the estimate of the PSF, we deconvolve the image and subsequently set the intensities in the occulted pixels in the deconvolved image again to zero, which yields a better approximation of the true image. Afterwards, we use this newly derived approximation to the true image together with the observed intensities of the occulted pixels in the original image to derive the next approximations of the PSF, and iterate until the approximation of the true image and the PSF both converge. In this approach, the accuracy of the final PSF depends only on the closeness of the final deconvolved image to the true image in the illuminated portion of the image plane. 

Finally, we explore the accuracy and speed of convergence of the PSF core and tail in more detail. The coefficients describing the tail of the PSF are related to long-distance scattered photons, i.e., they mostly originate from deep within the illuminated region where $I_\text{o}$ is almost exactly $I_\text{t}$ (see Equation \ref{eq:IoIt}). Therefore, the coefficients related to the derived PSF tail are expected to be highly reliable. For the core of the PSF, which depends on the quality of the estimates of the true intensities in the edge region, the situation is more complex. In the first iteration, the assumed true intensities in the edge region are underestimated, because the photons from the occulted region have not yet been redistributed to the illuminated part. As the true intensities are underestimated but the observed intensities are fixed, it follows from Equation~\ref{eq:psf} that the coefficients of the PSF core are initially overestimated. In the second iteration, this overestimation results in those photons scattered into the occulted region being redistributed into the illuminated region, particularly for those photons close to the edge region. This greatly improves the quality of the estimate of the true intensities, and consequently results in good estimates of the PSF coefficients by the second iteration. Subsequent iterations further adjust the PSF coefficients until the estimate of the true intensities in the edge region and the associated PSF coefficients both converge.

\subsection{Improving existing PSFs} \label{sec:revisingPSFs}

Existing or published PSFs can also potentially be improved using this methodology. Usually, the coefficients related to the core region are well known either from theoretical models or from observations of point sources, whereas the coefficients describing the PSF tail are difficult to fit. In our methodology, the coefficients describing the PSF tail depend mostly on the well-known true intensities deep within the illuminated region, and only weakly on the less constrained true intensities in the edge region. Thus, an adaption of our methodology enables one to   revise the tail coefficients of an otherwise well-known PSF. Let us denote the known PSF as $\overline{\text{psf}}$, and the missing portion of the PSF describing the tail as $\widetilde{\text{psf}}$. Since photons scattered far away, which are related to the PSF tail, have not been accounted for in the known PSF, the derived true intensities $I_\text{t,derived}$ of point sources were underestimated by the number of long-distance scattered photons, 
\begin{equation}
I_\text{t,derived} =  I_\text{t} \left(1 - \sum_{\pmb{\Delta r}>0} \widetilde{\text{psf}}_{\pmb{\Delta r}}\right) .
\end{equation}
Following Equation~\ref{eq:defpsf}, this underestimation of the derived true intensities combined with the fixed observed intensities resulted in an overestimation of the known PSF coefficients by the factor $\left(1 - \sum_{\pmb{\Delta r}} \widetilde{\text{psf}}_{\pmb{\Delta r}}\right)^{-1} $.
By discretizing the unknown PSF describing the tail into segments, this factor becomes $\left( 1 - \sum_S n_S\ \widetilde{\text{psf}}_S\right)^{-1}$, where $n_S$ is the number of pixels in a PSF segment. Combining the unknown PSF coefficients of the tail $\widetilde{\text{psf}}_S$ with the known PSF coefficients  $\overline{\text{psf}}_{\pmb{\Delta r}}$ corrected for their overestimation yields
\begin{equation}
I_\text{o,$\pmb{r}$} =   \sum_{S}  \widetilde{\text{psf}}_S \sum_{\pmb{\Delta r} \text{ in } S} I_\text{t,$\pmb{r}$-$\pmb{\Delta r}$}  + \left( 1 - \sum_S n_S\ \widetilde{\text{psf}}_S\right) \sum_{\pmb{\Delta r}}  I_\text{t,$\pmb{r}$-$\pmb{\Delta r}$} \ \overline{\text{psf}}_{\pmb{\Delta r}}. \label{eq:psf_revise}
\end{equation}
This equation can be fitted to the observed and true intensities analogous to Equation~\ref{eq:psf} to obtain the unknown tail coefficients, $\widetilde{\text{psf}}_S$. The final revised PSF coefficients are then  given by 
\begin{equation}
    \text{psf}_{\pmb{\Delta r}} = \widetilde{\text{psf}}_{S |_{\pmb{\Delta r} \text{ in } S}} +  \left( 1 - \sum_S n_S\ \widetilde{\text{psf}}_S\right) \ \overline{\text{psf}}_{\pmb{\Delta r}} .
\end{equation}
We note that the solution to this fit, i.e., the newly derived PSF coefficients $\widetilde{\text{psf}}_S$, is degenerate, i.e., that there are several solutions for the  $\widetilde{\text{psf}}_S$ that result in the same  composed PSF coefficients $\text{psf}_{\pmb{\Delta r}}$. This becomes clear when looking at an example where the known PSF is equivalent to the true PSF. In this case, there are two solutions for the newly derived PSF coefficients $\widetilde{\text{psf}}_S$: (1) the newly fitted PSF coefficients $\widetilde{\text{psf}}_S$ are all zero, and (2) the newly fitted PSF coefficients  $\widetilde{\text{psf}}_S$ are equivalent to the true PSF coefficients. Both fit solutions result in the same final assembled PSF coefficients $\text{psf}_{\pmb{\Delta r}}$. 

Furthermore, we note that our formulation preserves diffraction patterns and the structure of the PSF core explained by the known PSF. The newly derived true intensities increase by a factor of $ \left( 1 - \sum_S n_S\ \widetilde{\text{psf}}_S\right)^{-1}$ while the known PSF coefficients decrease by $\left( 1 - \sum_S n_S\ \widetilde{\text{psf}}_S\right)$. These terms cancel, and the predicted image intensities from the known PSF, which describes the PSF core and diffraction patterns around a point source, remain constant.

\subsection{Uniqueness of the solution} \label{sec:uniquness}

\begin{figure}
    \includegraphics[width=\textwidth]{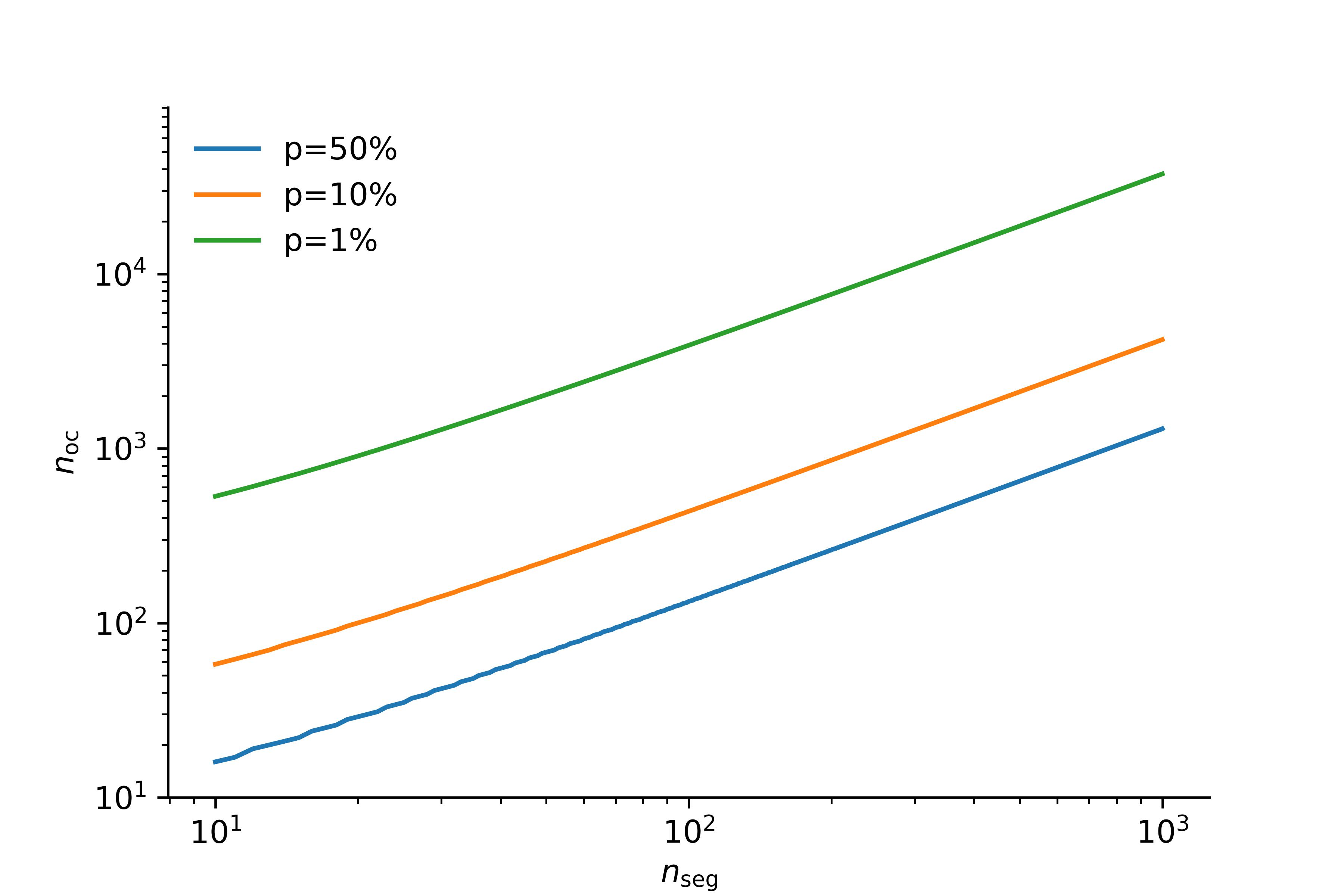}
    \caption{Required number of occulted pixels $n_\text{oc}$ for having a probability of more than \SI{99}{\percent} for a unique solution versus the number of PSF segments $n_\text{seg}$ to fit. The blue line assumes a probability of \SI{50}{\percent} that a single $a_{\pmb{r}}$ is non-negative, the orange line a probability of \SI{10}{\percent}, and the green line a probability of \SI{1}{\percent}.}
    \label{fig:probability}
\end{figure}

We now discuss if the methodology for deriving the instrumental PSF described in Section~\ref{sec:general} always converges to the correct solution, or, in other words, if the result of this procedure is unique. We are able to show that the solution is unique for a large class of occultation masks that are homogeneously illuminated by assuming that the solution is not unique and showing that this assumption results in a contradiction.

If the solution is not unique, then two different sets of assumed true images and PSFs, $\{ I_{1,\text{t}}, \text{psf}_1 \}$  and $\{I_{2,\text{t}},\text{psf}_2\} $, exist that result in this same observed image $I_{\text{o}}$. First, we focus on the uniquness of the true images. Without loss of generality, we assume that the true intensity in the illuminated portion of the image $I_{1,t} > I_{2,t}$. 
We define the $\Theta$ function to be zero if a pixel $\pmb{x}$ is in the occulted region and one if the pixel is in the illuminated region, 
\begin{equation}
    \Theta(\pmb{x}) = \begin{cases}
         0 \quad \text{for}\ I_\text{\{1,2\},t,\pmb{x}} = 0, \\
         1 \quad \text{for}\ I_\text{\{1,2\},t,\pmb{x}} > 0.
                \end{cases}
\end{equation}
We also define the location $\pmb{r}$ to be in the occulted region of the image.
Subtracting Equation~\ref{eq:psf} for the second true image, $I_{2,t}$ from Equation~\ref{eq:psf} for the first true image, $I_{1,t}$, yields
\begin{equation}
    0 \ =\ \left(  I_\text{1,t} \sum_{S} \sum_{(\pmb{\Delta r}>0) \text{ in } S} \Theta(\pmb{r}-\pmb{\Delta r})  \ \text{psf}_{1,S}   \right) \ -\ \left(  I_\text{2,t} \sum_{S} \sum_{(\pmb{\Delta r}>0) \text{ in } S} \Theta(\pmb{r}-\pmb{\Delta r})  \ \text{psf}_{2,S}  \right). \label{eq:defuniquness}  
\end{equation}
We have omitted the terms for $\pmb{\Delta r} = 0$, as in this case $\Theta(\pmb{r}) = 0$.

Since $I_\text{1,t} > I_\text{2,t}$, it follows that either
\begin{align}
    \sum_{S} \sum_{(\pmb{\Delta r}>0) \text{ in } S} \Theta(\pmb{r}-\pmb{\Delta r})  \ \text{psf}_{1,S} \quad &= \quad  \sum_{S} \sum_{(\pmb{\Delta r}>0) \text{ in } S}  \Theta(\pmb{r}-\pmb{\Delta r}) \ \text{psf}_{2,S} \quad  = 0 \label{eq:cond1} \\
\intertext{or}
   \sum_{S} \sum_{(\pmb{\Delta r}>0) \text{ in } S} a_{\pmb{r}}\  \Theta(\pmb{r}-\pmb{\Delta r})  \ \text{psf}_{1,S} \quad &< \quad  \sum_{S} \sum_{(\pmb{\Delta r}>0) \text{ in } S} a_{\pmb{r}}\   \Theta(\pmb{r}-\pmb{\Delta r}) \ \text{psf}_{2,S}.  \quad \quad \label{eq:cond2}
\end{align}
We have multiplied Condition~\ref{eq:cond2} with an arbitrary positive factor $a_{\pmb{r}}$, so that the  inequality keeps its sign, as a preparation for our subsequent discussion.
Condition~\ref{eq:cond1}~or~\ref{eq:cond2} have to be fulfilled for each occulted pixel $\pmb{r}$. Therefore, Condition~\ref{eq:cond1} defines a system of equations and  Condition~\ref{eq:cond2} a system of inequalities, where the number of occulted pixels $\pmb{r}$ determines the number of lines in the system of equations or system of inequalities, and the number of PSF coefficients $\text{psf}_S$ determines the number of variables.

We first lead Condition~\ref{eq:cond1} to a contradiction. As long as the system of equations of Condition~\ref{eq:cond1} is well-defined, i.e., as long as each PSF coefficient $\text{psf}_S$ with $\pmb{\Delta r}>0$ appears at least once in the system of equations and as long as we have at least as many occulted pixels as PSF coefficients, the solution of this system of equations is unique. It follows that the trivial solution 
\begin{equation}
    \text{psf}_{1,S_{|\pmb{\Delta r} >0}} = \text{psf}_{2,S_{|\pmb{\Delta r} >0}} = 0
\end{equation}
is the only solution. This equation infers that the PSF does not scatter any photons. However, there are scattered photons in the occulted region; therefore, Condition~\ref{eq:cond1} results in a contradiction.

The argument for Condition~\ref{eq:cond2} is more complex. To lead Condition~\ref{eq:cond2} to a contradiction, we aim at transforming Condition~\ref{eq:cond2} into
\begin{equation}
     \sum_{S} \sum_{(\pmb{\Delta r}>0) \text{ in } S} \text{psf}_{1,S} <  \sum_{S} \sum_{(\pmb{\Delta r}>0) \text{ in } S} \text{psf}_{2,S}. \label{eq:cond2_2}
\end{equation}
In Inequality~\ref{eq:cond2_2}, the summation  runs over the entire PSF except for the center pixel of the PSF. Therefore, Inequality~\ref{eq:cond2_2} infers that $\text{psf}_1$ scatters a smaller percentage of its intensity away than  $\text{psf}_2$.
A smaller scattering rate implies that the observed intensities are closer to the true intensities, i.e., that $I_{1,t} - I_{\text{o}} < I_{2,t} - I_{\text{o}}$, and it follows that $I_{1,t} < I_{2,t}$.
However, we have assumed $I_{1,t} > I_{2,t}$. Therefore, if the system of inequalities in Condition \ref{eq:cond2} can be transformed into Inequality~\ref{eq:cond2_2}, this will result in a contradiction. Then, neither Condition~\ref{eq:cond1} nor Condition~\ref{eq:cond2} can be fulfilled, and it would follow that the assumption $I_{1,t} > I_{2,t}$ was wrong and that consequentially  $I_{1,t} = I_{2,t}$, i.e., that the solution for the true image is unique.

Next, we investigate the likelihood for the existence of this transformation.  Condition \ref{eq:cond2} can be transformed to Inequality \ref{eq:cond2_2} by a linear combination of the inequalities in Condition \ref{eq:cond2}. The $a_{\pmb{r}}$ are the coefficients for this linear combination, and all the $a_{\pmb{r}}$ have to be non-negative so that Condition~\ref{eq:cond2_2}, as a sum of the Inequalities of Condition~\ref{eq:cond2}, does not change its sign. Therefore, we aim at finding a valid set $a_{\pmb{r}}$ that transforms Condition~\ref{eq:cond2} to Inequality~\ref{eq:cond2_2}. 
A sufficient condition is that a set $a_{\pmb{r}}$ exists that transforms each individual coefficient $\text{psf}_{S}$ from Condition~\ref{eq:cond2} to the corresponding one of Inequality~\ref{eq:cond2_2}. Summing all terms containing a given PSF coefficient $\text{psf}_S$ over $\pmb{r}$, i.e., over the lines of the system of inequalities in Condition \ref{eq:cond2},  and comparing it with Inequality~\ref{eq:cond2_2} defines the wanted transformation,
\begin{equation}
 \sum_{\pmb{r}} \sum_{(\pmb{\Delta r}>0) \text{ in } S} a_{\pmb{r}} \Theta(\pmb{r}-\pmb{\Delta r}) \text{psf}_S =  \sum_{(\pmb{\Delta r}>0) \text{ in } S} \text{psf}_S. \label{eq:ap_3}
\end{equation}
Equation~\ref{eq:ap_3} is a system of equations, where the PSF coefficients $\text{psf}_S$ define the lines and the $a_{\pmb{r}}$ are the variables.
We usually have significantly fewer PSF coefficient to fit than occulted pixels in the image. Therefore, we have significantly fewer lines in the System of Equations~\ref{eq:ap_3} than variables $a_{\pmb{r}}$, and consequentially System of Equation~\ref{eq:ap_3} is strongly underdetermined and the solution for the $a_{\pmb{r}}$ is degenerate.

In the following, we explore the probability that for at least one of these degenerate solutions all the $a_{\pmb{r}}$ are non-negative. Let us denote the probability that a single coefficient $a_{\pmb{r}}$ is non-negative as $p$, the number of occulted pixels as $n_\text{oc}$, the number of PSF segments to fit as $n_{\text{seg}}$, and the number of solutions as $n_\text{sol}$. 
A lower boundary for the number of solutions $n_\text{sol}$ can be estimated using combinatorics. Having $n_{\text{seg}}$ lines in the System of Equations~\ref{eq:ap_3}, we need exactly $n_{\text{seg}}$ variables, i.e., occulted pixels, to derive a single solution. Having $n_\text{oc}$ occulted pixels in total, the number of possibilities to draw $n_{\text{seg}}$ occulted pixels from $n_\text{oc}$ occulted pixels is 
\begin{equation}
    n_\text{sol} = \frac{n_\text{oc}!}{\left(n_\text{oc} - n_{\text{seg}}\right)!\ n_{\text{seg}}!}. \label{eq:combinations}
\end{equation}
For a single solution derived from $n_{\text{seg}}$ occulted pixels, the probability that all $a_{\pmb{r}}$  are non-negative is $p^{n_{\text{seg}}}$. The probability that not a single solution exists where all the $a_{\pmb{r}}$ are non-negative is $\left(1 - p^{n_{\text{seg}}}\right)^{n_\text{sol}}$. It follows that the probability $P$ that at least one solution exists where all $a_{\pmb{r}}$ are non-negative is
\begin{equation}
    P(p, n_\text{oc}, n_\text{sol}) = 1 - \left(1 - p^{n_{\text{seg}}}\right)^{n_\text{sol}}. \label{eq:probability}
\end{equation}
Combining Equation \ref{eq:combinations} and Equation \ref{eq:probability} enables one to derive a rough estimate of the probability that at least one solution exists where all the $a_{\pmb{r}}$ are non-negative, or, alternatively, to estimate how many occulted pixel in an image are required so that the probability is higher than a certain threshold. In Figure \ref{fig:probability}, we plot the required number of occulted pixels $n_\text{oc}$ so that the probability for the existence of a set of non-negative~$a_{\pmb{r}}$ is higher than \SI{99}{\percent} versus the number of PSF segments $n_{\text{seg}}$ to fit. The results are plotted for an assumed probability $p$ of~\SI{50}{\percent},~\SI{10}{\percent}, and~\SI{1}{\percent}. For $100$~PSF segments and an assumed probability of $p=\SI{50}{\percent}$, we find that at least $133$~occulted pixels are required. When we assume $p=\SI{1}{\percent}$, i.e., that the probability to obtain a non-negative $a_{\pmb{r}}$~is very small, $3908$~occulted pixels are required. As we usually have hundreds of thousands of occulted pixels in an image, it follows that it is highly probable that a set of non-negative~$a_{\pmb{r}}$ exists that solves Equation~\ref{eq:ap_3}, and, consequentially, that the solution for the true image will be unique.

Finally, we check if the solution for the PSF is unique. With $I_{1,t} = I_{2,t}$, Equation~\ref{eq:defuniquness} can be rewritten as 
\begin{equation}
     \sum_{S} \sum_{(\pmb{\Delta r}>0) \text{ in } S} \Theta(\pmb{r}-\pmb{\Delta r})  \ \text{psf}_{1,S}  = \sum_{S} \sum_{(\pmb{\Delta r}>0) \text{ in } S} \Theta(\pmb{r}-\pmb{\Delta r})  \ \text{psf}_{2,S}. \label{eq:cond3}
\end{equation}
Condition \ref{eq:cond3} is a system of equations with one line for each occulted pixel $\pmb{r}$. As long as this system of equations is well-defined, the solution for the variables $\text{psf}_S$ is unique. It follows that the trivial solution 
\begin{equation}
    \text{psf}_{1,S} \ = \ \text{psf}_{2,S}
\end{equation}
is the only solution. Thus, $\text{psf}_1 =\text{psf}_2$, and the solution for the PSF is unique.

We note that the argument in this section is valid as long as Condition~\ref{eq:cond2} can be transformed to Condition~\ref{eq:cond2_2}. Our requirement that the coefficients $a_{\pmb{r}}$ in Equation \ref{eq:ap_3} are non-negative is a stronger constraint than necessary. However, it guarantees that Condition~\ref{eq:cond2_2} has the same sign as Condition~\ref{eq:cond2} and thus that Equation~\ref{eq:ap_3} is a valid transformation. As Equation~\ref{eq:ap_3} is independent of the PSF coefficients, this makes the entire argument independent of the a priori unknown PSF coefficients. This enables one to verify, independently of the unknown PSF coefficients, if a given occultation mask together with a given discretization of the PSF guarantees a unique solution. The solution is unique if a set of non-negative $a_{\pmb{r}}$ exists which solves System of  Equations~\ref{eq:ap_3}.

There is one simple cases where the uniqueness of the PSF is guaranteed: Exactly one pixel in the center of the image is illuminated. Then, for each occulted pixel,in the summation over the illuminated pixels in Inequality~\ref{eq:cond2} exactly one $\Theta(\pmb{r} - \pmb{\Delta r})$ is one. Having only one PSF coefficient per line remaining in the system of inequalities enables one to rewrite the System of Inequalities~\ref{eq:cond2} as Inequality~\ref{eq:cond2_2}, guaranteeing the uniqueness.

\section{Implementation}  \label{sec:implementation}

\begin{figure}
    \includegraphics[width=\textwidth]{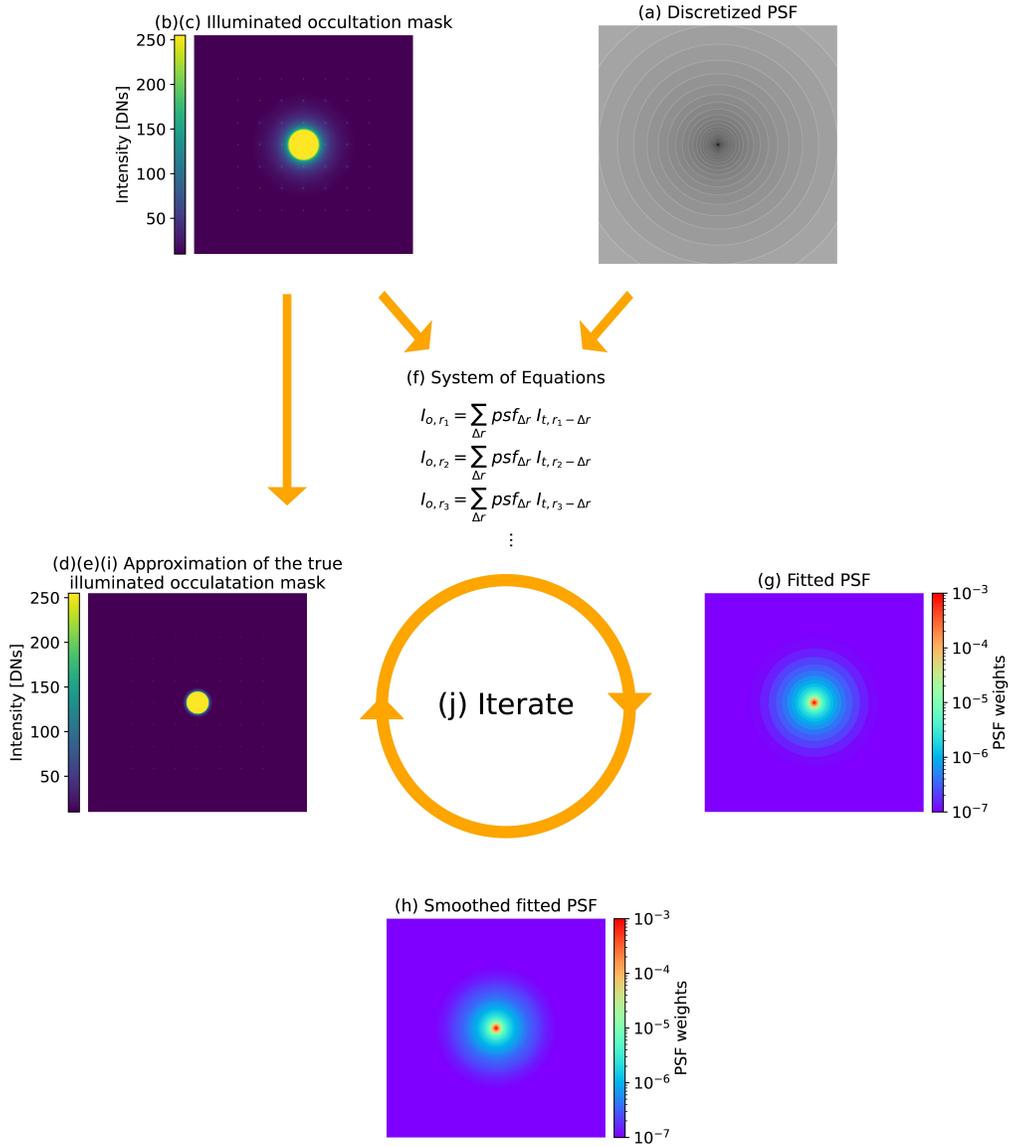}
    \caption{Schematic overview of the methodology. (a)~The PSF is discretized into segments. (b)~and~(c)~An occultation mask composed of a large hole and a grid of pinholes is illuminated from the back. Zoom in to see the pinholes. (d),~(e),~and~(i)~Occulted pixels are identified, and an approximation for the true illuminated occultation mask is derived. Zoom in to see the pinholes. (f)~The discretized PSF, the observed illuminated occultation mask, and the approximation of the true illuminated occultation mask are used to set up a system of equations. (g)~The system of equations is fitted to derive the PSF coefficients related to the PSF segments. (h)~The fitted PSF is smoothed to mitigate discretization effects. (i)~The smoothed fitted PSF is used to derive a revised approximation of the true illuminated occultation mask. (j)~Steps~(d)-(i) are iterated until the reconstructed image of the illuminated occultation mask and the fitted PSF both converge.}
    \label{fig:overview}
\end{figure}

In this section, we delineate the main stages for an implementation of our algorithm. 
The general procedure is as follows: (a)~discretize the PSF into segments, (b)~create an occultation mask, (c)~illuminate the occultation mask, (d)~identify the occulted pixels, (e)~derive an approximation for the true image of the illuminated occultation mask, (f)~set up the system of equations defined by Equation~\ref{eq:psf}, (g)~fit the PSF from this system of equations, (h)~postprocess the PSF, (i)~deconvolve the occulted image with the PSF to derive a revised approximation of the true image of the illuminated occultation mask, and (j)~repeat from steps (d)-(i) until the approximations of the true image and the PSF both converges. The final PSF can then be used to correct for instrumental effects of other images taken by the same imaging instrument.

In the following paragraphs, we will explain each of these steps in more detail. In Figure~\ref{fig:overview}, we visualize these steps by showing a schematic of the procedure; each panel has the label of the corresponding step or steps in our procedure.

\paragraph{(a) Discretizing the PSF}
First, we discretize the two-dimensional PSF spatially into segments (see Fig.~\ref{fig:overview}(a)). Each segment  corresponds to one PSF coefficient to be fitted. As the values of the PSF coefficients in the core of the PSF usually decreases rapidly within a few pixels from the peak, the core of the PSF should be determined at the full resolution of the detector. The PSF coefficients in the tail usually vary much more slowly and larger regions in the PSF tail can be aggregated to a single PSF coefficient. This aggregation improves the signal-to-noise (S/N) level when fitting the PSF coefficients later on and reduces the required computation time. A priori knowledge, such as diffraction patterns, can also be included, in which case each segment in the diffraction pattern should be assigned a PSF coefficient to be fitted.

\paragraph{(b) Creating occultation masks}
Our algorithm works with all kinds of partially occulted images. In laboratory settings, one can create various occultation masks that are specialized for different tasks. To resolve the core of the PSF, we recommend using an occultation mask containing several pinholes. Setting the pinhole diameters to a size of several pixels enhances the approximation of the true intensities in the pinholes and thereby the overall accuracy of our algorithm. 
If a diffraction pattern is apparent, the distance between the pinholes should be large enough so that the different diffraction patterns are distinguishable. Larger pinhole sizes are preferred as it results in more intensity in the diffraction pattern. To fit the tail of the PSF accurately, we suggest an occultation mask containing a single large hole, which drastically increases the total number of photons available. The concepts of these occultation masks can also be combined into a single occultation mask for convenience (see Fig.~\ref{fig:overview}(b)). 

\paragraph{(c) Illuminating the occultation mask}
Next, we uniformly illuminate a screen, set the occultation mask between the screen and the imaging device, and record the resulting partially occulted image (see Fig.~\ref{fig:overview}(c)).

\paragraph{(d) Identifying occulted pixels}
Pixels at the edge of the illuminated area will have photons scattered into the occulted area. However, there are no photons scattered back from the occulted area to the illuminated edge to counterbalance the photon loss. Consequently, the intensities decrease across the edge between the illuminated area and occulted area. As most photons are scattered over short distances and as the total numbers of photons is conserved, the edge can be estimated to be roughly at the \SI{50}{\percent} level of the illuminated edge intensities. We recommend defining fully occulted pixels to be at least one pixel away from the \SI{50}{\percent} intensity level of the edge to avoid partially illuminated pixels.

\paragraph{(e) Deriving an approximation for the true image of the illuminated occulation mask}
In the first iteration of our algorithm, we use the observed image of the illuminated occultation mask as a basis for the approximation of the true image; in subsequent iterations, the basis is the observed image deconvolved with the fitted PSF from the previous iteration.
There are two ways to derive the approximation for the true image from the basis. A general method is to set the intensities of the fully occulted pixels to zero, while keeping the intensities in the partially and fully illuminated pixels in the basis unchanged. In this approach, the true intensities in the partially and fully illuminated pixels are reconstructed by the deconvolution of the observed image with the fitted PSF from the previous iteration. Alternatively, in use cases where the positions of the occulted pixels are precisely known, as in certain laboratory settings, one may set the intensities in the fully occulted pixels to zero, set the intensities in the illuminated pixels to the average intensity in the illuminated region, and use a subpixel edge reconstruction to correctly determine the intensities in the partially illuminated pixels (see ref. \cite{nakashima2003}). In the remainder of this study, we use the general method to approximate the true image (Fig.~\ref{fig:overview}(e)).

\paragraph{(f) Setting up the system of equations}
Each fully occulted pixel intensity $I_{\pmb{r}}$ gives one line in the system of equations defined by Equation~\ref{eq:psf} (see Fig.~\ref{fig:overview}(f)). For fully each occulted pixel, the integrated intensities $I_\text{t,$\pmb{r}$-$\pmb{\Delta r}$}$ in the illuminated regions of the presumed true image associated with each PSF segment $\text{psf}_S$  have to be derived (see Equation~\ref{eq:psf}). Therefore, the computational cost scales as $n_\text{oc} n_\text{il}$, where $n_\text{il}$ is the number of illuminated pixels in the true image. With $n_\text{oc}$ and $n_\text{il}$ both being typically at the order of \num{e6}, the computational costs can become expansive. To reduce the computation time, for occulted pixels that are far enough from the illuminated edge, we recommend grouping several contiguous occulted pixels into a superpixel that has their average intensity. This merging also increases the S/N in the observed intensity of the occulted superpixel. We note that the system of equations to be solved is not limited to using a single image. Multiple images involving possibly different occultation masks can also be used to set up a combined system of equations. This enables one to determine the PSF by using several specialized occultation masks, i.e., one mask for deriving the PSF core, one mask for fitting the diffraction patterns, and one mask to determine the PSF tail.

\paragraph{(g) Fit the PSF}
Having set up the system of equations, we derive a multi-linear fit to the system of equations using the Levenberg-Marquardt algorithm \cite{levenberg1944, marquardt1963} to determine the PSF coefficients (see Fig.~\ref{fig:overview}(g)). To optimize the fitting process, the PSF coefficients, which define the columns of the system of equations, should be constrained to allow for only non-negative values. Furthermore, as the observed intensities in the occulted pixels can be expected to vary over several orders of magnitudes from the illuminated edge to far within the occulted region, each line of the system of equations should be normalized by its observed intensity. 

There are several procedures by which this system of equations can be fit:
\begin{enumerate}
    \item  When the number of PSF coefficients is small, one can perform a simple multi-linear fit to determine the weights in the PSF segments. 
    \item  When the number of PSF coefficients is large, one has to prevent the fitting algorithm from terminating in a local minimum. The easiest way to do this is to start the fitting process at a low resolution, i.e., a small number of coefficients to be fitted, and to iteratively increase the resolution. For this purpose, we recommend placing a low-resolution adaptive grid on top of the discretized PSF. Each node in the adaptive grid becomes a support coefficient. The PSF coefficients are linked to the support coefficients by a spline interpolation. At each iteration, we fit the support coefficients to the system of equations. Subsequently, we increase the resolution of the adaptive grid and iterate until the full resolution given by the PSF discretization is reached. This guarantees a reasonable initialization of the fit coefficients at each iteration, speeds up the fitting process, and reduces the risk of terminating in a local minimum. 
    \item When the number of PSF coefficients is very large, one can decompose the fitting problem into several smaller fitting problems. Occulted pixels far from the illuminated edge only receive long-distance scattered photons, which are related to PSF coefficients far from the PSF center. Therefore, one can first select lines in the system of equations related to occulted pixels far from the illuminated edge and fit the associated PSF coefficients. Then, one can consecutively select lines of the system of equations that are related to occulted pixels closer to the illuminated edge to fit PSF coefficient closer to the PSF core, while keeping the already fitted PSF coefficients constant. The reduced number of data points and fit coefficients within each subsystem of equations thereby greatly speeds up the entire fitting process.
    \item When noise is present, one has to avoid overfitting, i.e., fitting the specific solution of the noise-dependent image instead of a general solution. There are two ways to circumvent overfitting: First, instead of using all lines of the system of equations at once, one can use a random subset, fit the PSF coefficients, and repeat this procedure many times. The final PSF coefficients are then given as the mean of the individual fits. Second, one can add a regularization parameter to the fitting process, such as the Ridge regularization~\cite{Tikhonov1943} or Lasso regularization~\cite{Tibshirani1996}. Then, each column of the system of equations has to divided (i.e., normalized) by the estimated size of the associated PSF coefficient beforehand, so that the regularization works with the same strength on all coefficients. After having obtained the final fit coefficients, the normalization of the fit coefficients has to be removed.
    \item When revising the coefficients of a known PSF by Equation~\ref{eq:psf_revise}, the solution for the fitted missing PSF coefficients is degenerate. In general, one is interested in the solution that least modifies the known PSF, i.e., the solution where the fit coefficients describing the missing portion of the PSF are minimal. This can be achieved by adding the sum of the fit coefficients describing the missing portion of the PSF as a regularization parameter to the fit.
    \item When the analytical form of the PSF can be guessed, e.g., from a previous fit of the PSF or from theoretical work, one can re-parameterize the PSF coefficients by this analytical expression as an initial step. This enables one to fit for the free parameters in the analytical expression of the theoretical PSF. 
\end{enumerate}

Finally, we note that the system of equations is usually strongly overdetermined, as there are typically more occulted pixels than PSF coefficients to determine. Using only a subset of the system of equations is typically enough to reliably constrain the fit and to greatly speed up the calculation. For this, the occulted pixels in this subset need to be evenly distributed over the entire occulted region, as pixels close to the illuminated edge mostly affect the quality of the core of the fitted PSF, while those far away constrain the tail of the PSF.

\paragraph{(h) Postprocessing the PSF}
Postprocessing the PSF is not strictly required as its effect on the reconstruction of true images is usually negligible. But it enables one to derive a continuous approximation of the discretized PSF. For this task, we recommend performing a large number of iterations of Laplacian smoothing on the PSF, where the weights within each segment have to be renormalized after each iteration to match the fitted weight. The larger the sizes of the PSF segments are, the more iterations are needed to derive the a good continuous approximation. This guarantees a smooth transition between the PSF segments while keeping the weights of the PSF segments correct (see Fig.~\ref{fig:overview}(h)). 

\paragraph{(i) Updating the approximation of the true image}
Next, the original image has to be deconvolved with the derived PSF to acquire an improved approximation for the true image (see Fig.~\ref{fig:overview}(i)). The deconvolution algorithm used has to retain the total image intensity as well as sharp edges, making the Richardson-Lucy algorithm a good choice \cite{Lee2011}. 
We note that the image deconvolution involved in this step is an ill-posed problem and limits one to the accuracy of the deconvolution algorithm chosen.  Therefore, even for the fully converged PSF solution, deviations from the true PSF can be expected. These deviations, however, are usually small. The overall accuracy will be further investigated in Section~\ref{sec:evaluation}.

\paragraph{(j) Iterating}
Finally, we repeat Steps (d)-(i) of this entire procedure until the true image and the PSF both converge.

\ \\[.1cm]
Having derived the instrumental PSF, we are able to reconstruct additional true images by deconvolving the associated observed images with the fitted instrumental PSF.

\section{Numerical experiments}  \label{sec:evaluation}

\begin{table}
    \centering
    \resizebox{\textwidth}{!}{%
    \begin{tabularx}{1.4\textwidth}{l|c c c c c c}
                             %& \multicolumn{4}{c}{PSF} \\
       Parameter\ / \ PSF             &   \begin{tabular}{c} Gaussian\\+ Lorentzian \\ (Section~\ref{sec:showcase1})\end{tabular}& \begin{tabular}{c} Gaussian\\+ Lorentzian \\ (Section~\ref{sec:showcase3})\end{tabular}&  \begin{tabular}{c} Elliptical \\ (Section~\ref{sec:showcase3})\end{tabular} & \begin{tabular}{c} Airy pattern \\+ coma and astigmatism\\(Section~\ref{sec:showcase3})\end{tabular} &
       \begin{tabular}{c} Diffraction pattern  \\ (Section~\ref{sec:showcase3})\end{tabular} & \begin{tabular}{c} AIA \\ (Section~\ref{sec:showcase3})\end{tabular} \\ \hline 
       Image: & & & & \\
        \quad Image resolution &  $1024\times 1024$   &  $1024\times 1024 $ & $1024\times 1024$                                          &  $1024\times 1024$ &  $1024\times 1024$    & $1024\times 1024$ \\
        \quad Occultation mask  & \begin{tabular}{c} 1 large hole \\ 48 pinholes \end{tabular} & 1 large hole & 1 large hole & \begin{tabular}{c} 1 large hole \\ 48 pinholes \end{tabular} & \begin{tabular}{c} 1 large hole \\ 48 pinholes \end{tabular} & Partial solar eclipse \\
        \quad Occulted pixels & \SI{99}{\percent} & \SI{99}{\percent}  & \SI{99}{\percent} & \SI{99}{\percent} & \SI{99}{\percent} & \SI{23}{\percent}\\
       Fitting: & & & & \\
       \quad Fit function   & Multilinear & \begin{tabular}{c} Gaussian\\+ Lorentzian \end{tabular} & \begin{tabular}{c} Multilinear \\ + Support grid \\ \end{tabular} & \begin{tabular}{c} Multilinear \\ Subsystems of equations \\ \end{tabular} & Multilinear & Multilinear \\
       \quad \# Iterations      & 100 & 5 & 5 & 5 & 5 & 5 \\
       \quad \# PSF segments     & 40 & 200 & 480 & 480 & 80 & 50\\
       \quad \# Coefficients to fit  & 40 & 5  & 480 & 480 & 80 & 50 \\
       \quad \# Lines in SoE    & \num{1000} &\num{1000} &\num{3000} &\num{3000} &\num{1000} &  \num{1000} \\
       \quad \# Fit repetitions      &  1    &   1  & 3 & 10                    &    1    & 10   \\
       Accuracy: & & & & \\
       \quad MAPE &           \SI{1.3}{\percent} & \SI{0.4}{\percent}& \SI{4.3}{\percent}& \SI{2.5}{\percent} &\SI{1.4}{\percent} & Unknown\\
       \quad MAPE (center) & \SI{0.2}{\percent}&   \SI{0.2}{\percent} & \SI{0.3}{\percent}& \SI{1.2}{\percent}& \SI{0.3}{\percent} & Unknown \\
       \quad MAPE (core) & \SI{0.7}{\percent}&   \SI{0.4}{\percent} & \SI{4.6}{\percent}& \SI{2.8}{\percent}& \SI{1.0}{\percent} & Unknown\\
       \quad MAPE (tail) & \SI{1.7}{\percent}&    \SI{0.5}{\percent} & \SI{3.0}{\percent}& \SI{1.9}{\percent}& \SI{1.5}{\percent} & Unknown\\
       Runtime per iteration: & & & & \\
       \quad Setting up SoE   & \SI{8}{s} & \SI{9}{s} & \SI{14}{s} & \SI{14}{s} & \SI{8}{s} & \SI{3}{s} \\
       \quad Fitting SoE  & \SI{1}{s} & \SI{2}{s} & \SI{216}{s} & \SI{28}{s} & \SI{2}{s}& \SI{6}{s}\\
       \quad Deconvolution & \SI{1}{s} & \SI{1}{s} & \SI{1}{s} & \SI{1}{s} & \SI{1}{s} & \SI{1}{s} \\
       \quad Postprocessing & \SI{2}{s} & \SI{3}{s} & \SI{3}{s} & \SI{3}{s} & \SI{2}{s} & \SI{3}{s} \\
       \quad Overhead & \SI{3}{s} & \SI{3}{s} & \SI{8}{s} & \SI{8}{s} & \SI{3}{s} & \SI{3}{s}\\
       \quad Total & \SI{15}{s} & \SI{18}{s} & \SI{242}{s}  & \SI{54}{s} & \SI{16}{s} & \SI{16}{s}
    \end{tabularx}
    }
    \caption{Setups of our numerical experiments. SoE refers to System of Equations.}
    \label{table:timings}
\end{table}

In this section, we study the stability and accuracy of our algorithm. For five test cases, we convolve a numerical true image of an illuminated occultation mask with a true PSF to derive an observed illuminated occultation mask. Afterwards, we apply our algorithm on this observed illuminated occultation mask to reconstruct the PSF, and analyze the accuracy by comparing the reconstructed PSF with the true PSF. In Section~\ref{sec:showcase1}, we investigate in detail the accuracy of our algorithm for a simple cylindrically symmetric PSF as a proof of concept, and in Section~\ref{sec:showcase2}, we analyze the noise stability of our algorithm on this PSF. In Section~\ref{sec:showcase3}, we show reconstruction results for a cylindrically symmetric PSF where the functional form is known and only free parameters have to be determined, for an elliptical PSF, for a PSF consisting of an Airy pattern with coma and astigmatism aberrations, for a PSF that contains a diffraction pattern, and for the revised PSF of AIA \cite{lemen2012, pesnell2012}.

These test cases all have an image sizes of $1024\times 1024$ pixels. The tests were run using an NVIDIA GeForce RTX~2080 Ti graphics processing unit, i.e., a consumer graphics card, to set up the system of equations, for deconvolving the images, and for postprocessing the images, and on an AMD~Ryzen~9~3950X processor for fitting the system of equations. Table~\ref{table:timings} lists the runtimes and the details of the configurations. Typical runtimes for our algorithm are \SI{15}{seconds} to \SI{4}{minutes} per iteration, with $\le 5$ iterations required for convergence.

\subsection{Proof of concept} \label{sec:showcase1}

\begin{figure}
    \includegraphics[width=\textwidth]{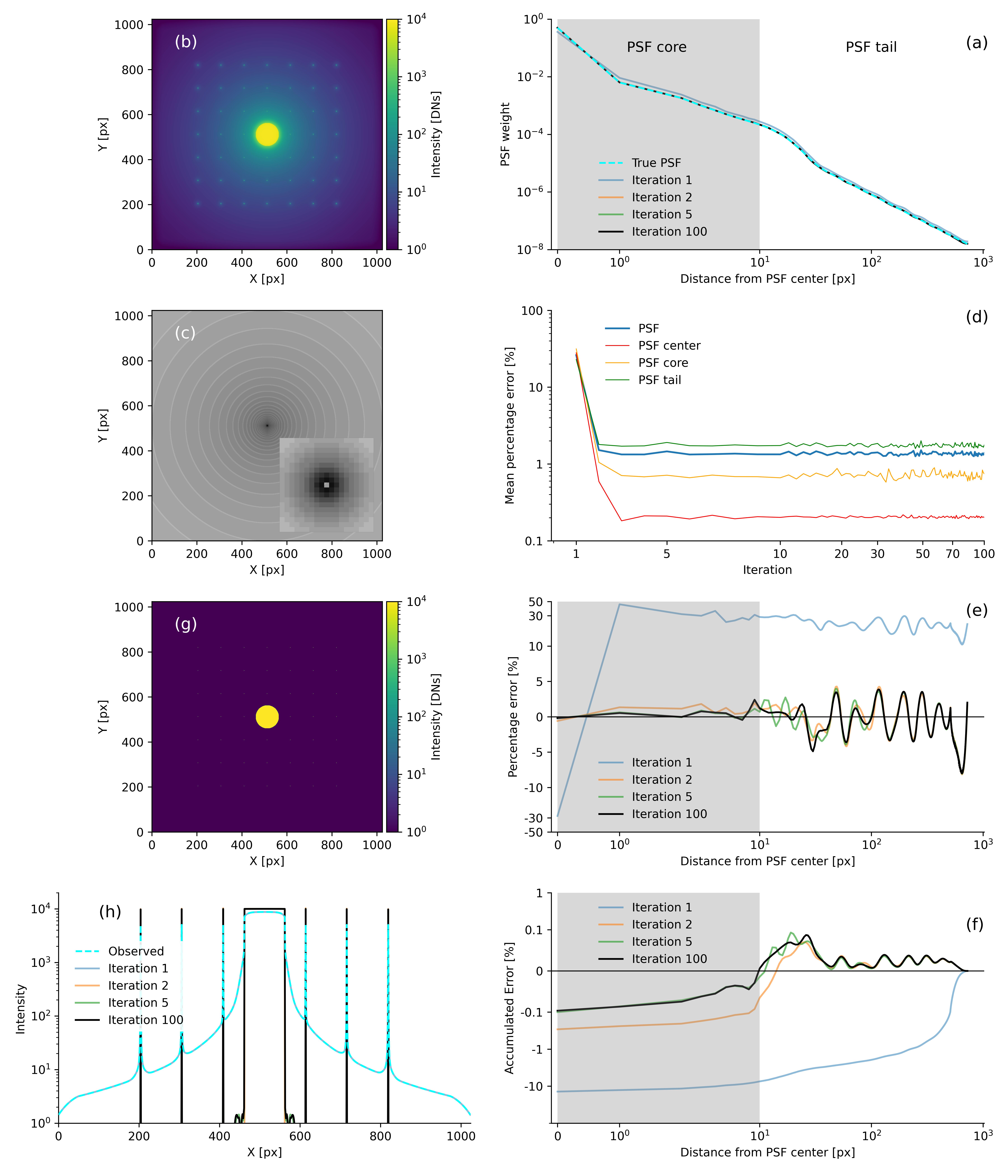}
    \caption{First test case. (a)~True and fitted weights of the PSF. From the second iteration on, the fitted weights are indistinguishable from the true PSF. (b)~Illuminated occultation mask. (c)~Discretized PSF; the inset with a size of $20\times 20$ pixels shows the discretization of PSF core region. Each gray level corresponds to one PSF coefficient of the fit. (d)~Evolution of the convergence of the fit. (e)~Deviations of the fitted PSF weights to the true PSF weights. (f)~Accumulated error of the PSF weights up to a given distance to the PSF center. (g)~Illuminated occultation mask deconvolved with the final PSF. Zoom in to see the pin holes. (h)~Intensities along a horizontal slice through the center of the deconvolved illuminated occultation mask.}
    \label{fig:converges}
\end{figure}

As a first test case, which is presented in Figure \ref{fig:converges}, we define a cylindrically symmetric PSF composed of a Gaussian core with a $\sigma = 10~\text{pixels}$ and a Lorentzian tail,
\begin{equation}
    \text{psf}(r) = \num{2.4e-4}\ \exp{\left(-\frac{r^2}{200}\right)} + \frac{0.5}{1 + 61 r^2}, \label{eq:cylindricalPSF}
\end{equation}
where $r$~is the distance to the PSF center (Fig.~\ref{fig:converges}(a)). This PSF scatters about~\SI{50}{\percent} of the photons over the image plane. 

We use an occultation mask that contains one large hole with a radius of 50~pixels in its center and 48~pinholes each with a size of 1~pixel distributed over the occultation mask. The true image of the illuminated occultation mask is the occultation mask with the intensities in each pixel within the holes set to \num{10000}~digitial numbers~(DNs). The observed image of the illuminated occultation mask is the true image convolved with the PSF. The observed image is shown in Figure~\ref{fig:converges}(b); each of the holes exhibits a halo due to the scattered light.

To reconstruct the PSF from the observed illuminated occultation mask, we discretize the PSF into shells. We define the core region of the PSF as the area within 10~pixels around the PSF center and the tail region as the remaining PSF area. We describe the core region of the PSF with 10~concentric shells, each having a width of $0.5$~pixels, and then 5~additional concentric shells each with a width of $1$~pixel. The PSF tail region is segmented into 25~concentric shells in logarithmic steps. This results in a total of 40~PSF coefficients to determine (Fig.~\ref{fig:converges}(c)). Next, we approximate the true illuminated occultation mask by setting the intensities in the occulted pixels to~\SI{0}{DNs}. We set up System of Equations~\ref{eq:psf}, randomly draw $10^3$~lines from the approximately $10^6$~lines of the System of Equations, and fit the PSF coefficients by a simple multi-linear fit. We assemble the PSF using the fitted PSF coefficients and derive a continuous approximation of the PSF by apply $10^4$~iterations of Laplacian smoothing. This large number of iterations of Laplacian smoothing accounts for the large size of the PSF segments in our PSF tail. Finally, we deconvolve the observed illuminated occultation mask with this smoothed PSF to derive the next approximation of the true illuminated occultation mask. To be able to study the rate and quality of convergence of our algorithm, we iterate this procedure $100$~times, and arrive at the final PSF.

 We now analyze the accuracy of the reconstructed PSF. We show the fitted PSF at the first, second, fifth, and 100th iteration together with the true PSF in Figure~\ref{fig:converges}(a). In the first iteration, the PSF is overestimated, but from the second iteration on, the fitted PSF shows excellent agreement with the overall shape of the true PSF over all eight order of magnitudes. The mean of the absolute percentage error (MAPE) of the fitted to the true PSF coefficients is shown in Figure~\ref{fig:converges}(d). There, we plot the MAPE of the PSF center coefficient, of the PSF core region, of the PSF tail region and of the entire PSF versus the iteration number. By the second iteration, the MAPE of the entire PSF decreases to \SI{1.5}{\percent}, and by the third iteration, the MAPE can be considered to have converged to \SI{1.3}{\percent}.  The MAPE of the PSF center coefficient converges to \SI{0.2}{\percent}, where we remind the reader that the PSF center coefficient determines the percentage of photons that are not scattered. The MAPE of the PSF core region converges to \SI{0.7}{\percent}, and the MAPE of the PSF tail region to \SI{1.7}{\percent}. These values show that all the PSF segments are fitted accurately over all eight orders of magnitude. 
 
In Figure~\ref{fig:converges}(e), we show the percentage error for the PSF weights along the PSF cross section. Starting from the second iteration, the percentage errors in the PSF core region are below \SI{2}{\percent}, while the percentage errors in the tail region oscillate around the true solution with an amplitude of \SI{\approx 3}{\percent}. The larger amplitude in the last oscillation at the very end of the tail is an effect of the discretization of the PSF. As there is no further outer shell, the boundary condition for the smoothing is not well defined, resulting in larger maximum deviations for the outermost shell. These oscillations mostly arise from the limited accuracy of the solver for the system of equations. As the absolute errors in the PSF weights due to these oscillations are only in the range of \numrange{e-4}{e-9} and because the absolute errors decrease with increasing distances from the PSF center, these oscillations have almost no effect on real-world applications. A more relevant parameter for real-world applications is the accumulated error, i.e., the total signed error of the PSF weights from the PSF center up to a given distance to the PSF center. This corresponds to the absolute error in the number of photons that are scattered up to that given distance, and is shown in Figure~\ref{fig:converges}(f). For each distance, the absolute value of the accumulated error is always smaller than \SI{0.1}{\percent}, which shows the high fidelity in the spatial distribution of the scattered photons.

Next, we evaluate the quality of the reconstructed image. In Figure~\ref{fig:converges}(g), we show the reconstructed image, i.e, the observed illuminated occultation mask deconvolved with the final PSF. The intensity in the large hole is homogeneous without showing a halo, and the 48~pinholes appear as point sources with a size of one pixel. The reconstructed intensities along a horizontal slice through the center of the image are plotted in Figure~\ref{fig:converges}(h) for the first, second, fifth, and final iteration of our algorithm. The intensities in the occulted areas along the slice were originally between~\num{1.3} and \SI{1700}{DNs} in the observed image (cyan line), but they become negligible from the second iteration onward (orange, green, and black line; the orange and green lines are mostly covered by the black line). The peak intensities in the pinholes were \SI{\approx 5000}{DNs} in the observed image, compared to \SI{10000}{DNs} in the true image. Deconvolving the observed image with the fitted PSF increases the peak intensities to between~\num{9650} and \SI{9950}{DNs}. The same effect is apparent in the large hole. The observed intensities ranged from \SIrange{7000}{8750}{DNs} for the pixels within the large hole, compared to \SI{10000}{DNs} in the true image. Deconvolving the observed image increases the intensities to between~\num{9975} and \SI{10006}{DNs}. Ringing in the reconstructed image due to the slight oscillations in the fitted PSF is not visible as the amplitude of the ringing is much smaller than \SI{1}{DN}. Finally, we focus on the intensity drop at the illuminated edges. At the edge of the pinholes in the deconvolved image, the intensities drop from \SI{9950}{DNs} to \SI{<0.05}{DNs} within one pixel. At the edge of the large hole in the deconvolved image, the intensities drop from \SI{9975}{DNs} to \SI{2}{DN} within one pixel. Thus, we find excellent reconstructions of the pinholes, the large hole, and the edge of the illuminated areas.

\subsection{Noise stability} \label{sec:showcase2}

\begin{figure}
    \includegraphics[width=\textwidth]{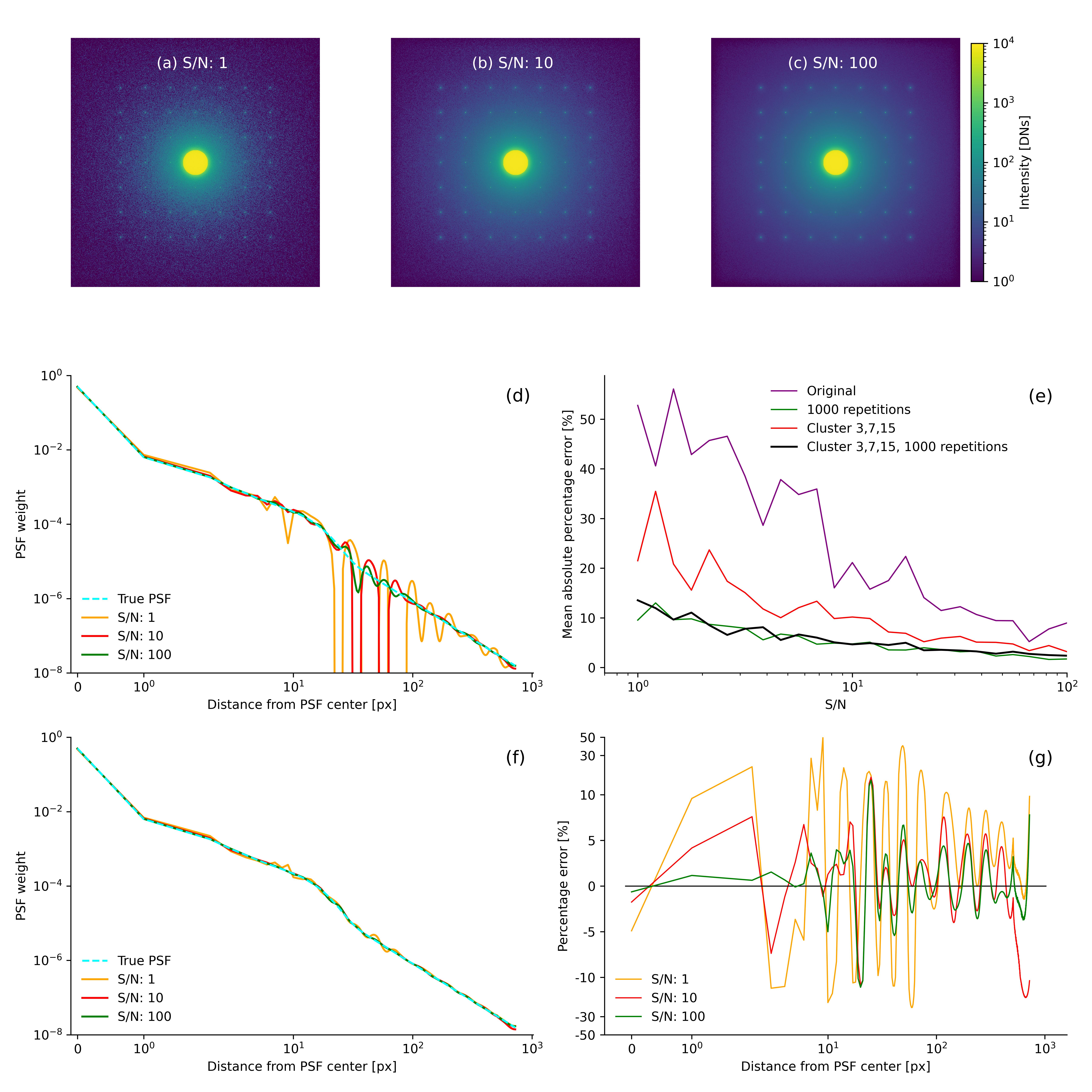}
    \caption{Robustness against image noise. (a)-(c)~Observed illuminated occultation masks for S/N~levels of~1,~10, and~100, respectively. (d)~Fitted PSFs for these noise levels without applying noise mitigation techniques. (e)~Evolution of the MAPEs of the fitted to the true PSF coefficients with increasing the S/N~level, derived for the labelled noise mitigation techniques. (f)~Fitted PSF for S/N~levels of~$1$,~$10$, and~$100$, where the combined clustering and fit repetition noise mitigation technique was applied. (g)~Associated deviations of the fitted PSF weights to the true PSF weights for these S/N~levels.}
    \label{fig:noise}
\end{figure}

In our methodology, the PSF is fitted from the intensities in the approximated true image and the observed intensities in the occulted regions. In the occulted regions, the signal is in general low, and consequentially the S/N can be low. Here, we analyse the effect of image noise on the quality of the fitted PSF, and test two methodologies for mitigating image noise: clustering pixels into a superpixel and using multiple fit repetitions.  

Image noise can affect the quality of the fitted PSF core, the fitted PSF tail, or both. In the PSF core region, the PSF weights typically fall off rapidly, and the most important occulted pixels for fitting these weights are occulted pixels close to the illuminated edge. As the intensities in the occulted region close to the illuminated edge decreases rapidly, these occulted pixels cannot be clustered without affecting the quality of the reconstructed PSF. Therefore, to improve the quality of the PSF core, averaging multiple fit repetitions is the preferred technique. For the PSF tail region, the most important occulted pixels for fitting these weights are those at locations far from the illuminated edge. Far from the illuminated pixels, the observed intensity decreases slowly, and therefore many of these occulted pixels can be clustered to improve the S/N in these pixels. Hence, to improve the quality of the PSF tail, clustering is the preferred technique. 

In the following, we simulate the effect of image noise and investigate its effect on the quality of the reconstructed PSF for the PSF and occultation mask of Section~\ref{sec:showcase1}. 
The average observed intensity in the occulted region is \SI{14}{DNs}, with a maximum intensity of \SI{1700}{DNs} close to the edge to the illuminated region and a minimum intensity of \SI{1.3}{DNs} at the edges of the detector. Therefore, when a constant noise level is present, the S/N varies greatly from the edge of the illuminated region to the edge of the detector. Here, we define the S/N~level as the average intensity in the occulted image region over the noise level. Assuming S/N~levels of~$1$,~$10$, and~$100$, the corresponding Poisson noise levels are \SI{14}{DNs}, \SI{1.4}{DNs}, and \SI{0.14}{DNs}. 

For each S/N~level, we select the intensities in the observed image from the associated Poisson distributions to create the noisy images shown in (Fig~\ref{fig:noise}(a)-(c)) and subsequently re-derive the PSF. Using these images, we test two noise mitigating techniques: (1)~we cluster $3\times 3$~occulted pixels into a superpixel when the minimum distance to the illuminated pixels is at least 15~pixels, $7\times 7$~occulted pixels when the minimum distance is at least 35~pixels, and $15\times 15$~occulted pixels when the minimum distance is at least 75~pixels. The minimum distance from the illuminated edge chosen corresponds to ten times half the edge length of the superpixel. This choice satisfies the condition that the intensities in the clustered pixels have to vary slowly across the superpixels. And~(2), we repeat the fitting procedure $1000$~times and derive the final fit as the average of the individual fits. As a baseline for the noise mitigation techniques, we use a simple multi-linear fit without noise mitigation. For maximum noise mitigation, we combine the cluster technique with the fit repetition technique. 
 
In Figure~\ref{fig:noise}(d), we show the fitted PSFs for S/N~levels of~$1$,~$10$, and~$100$ for the baseline configuration. At a S/N~level of~$1$, the fitted PSF shows strong oscillations around the true PSF. The amplitude of the oscillations decreases with increasing S/N~level. The evolution of the MAPE between the fitted and true PSF coefficients with increasing S/N~level is shown in Figure~\ref{fig:noise}(e). At a S/N~level of~$1$ the MAPE is~\SI{53}{\percent}; it decreases to~\SI{21}{\percent} at a S/N~level of~$10$ and to~\SI{9}{\percent} at a S/N~level of~$100$. When applying the clustering technique alone, the MAPE is~\SI{21}{\percent} for a S/N of~$1$, \SI{10}{\percent} for a  S/N of~$10$, and~\SI{3}{\percent} for a S/N of~$100$. When applying the fit repetition technique alone, the MAPE is~\SI{10}{\percent} for a S/N of~$1$, ~\SI{5}{\percent} for a S/N of~$10$, and~\SI{2}{\percent} for a S/N of~$100$. The MAPEs for the combined clustering and fit repetition configuration performs about as well as the fit repetition technique alone.  For comparison, the MAPE without noise is \SI{1.3}{\percent} (see Section~\ref{sec:showcase1}).

In Figure~\ref{fig:noise}(f), we plot the fitted PSFs for S/N~levels of~$1$,~$10$, and~$100$ for the combined clustering and fit repetition configuration. For a S/N~level of $1$, only moderate oscillations are apparent, and for S/N~levels of~$10$ and~$100$, the oscillations around the true solution almost vanish. In Figure~\ref{fig:noise}(g), we show the associated deviations of the fitted PSF weights from the true PSF weights along a cross-section through the PSF center. For a S/N~level of~$1$, the fitted PSF oscillates with an average amplitude of \SI{20}{\percent} around the true PSF, for a S/N of~$10$, the average amplitude of the oscillations is about \SI{5}{\percent}, and for a S/N of~$100$, the average amplitude of the oscillations is about \SI{4}{\percent}. For comparison, the amplitude of the oscillations without noise is \SI{3}{\percent} (see Section~\ref{sec:showcase1}).

These cases show that noise mitigation techniques can successfully be applied to generate good fits even for mediocre S/N~levels. For cases where the S/N~level is too low for generating good results even after applying noise mitigation techniques, we recommend improving the S/N~level by either increasing the exposure time of the image or by enlarging the size of the holes to increase the total numbers of photons available. 
In general, we recommend good S/N~levels in the occulted region; the higher the S/N, the more accurate the reconstructed PSF will be.

\subsection{Further examples} \label{sec:showcase3}

\begin{figure}
    \includegraphics[width=\textwidth]{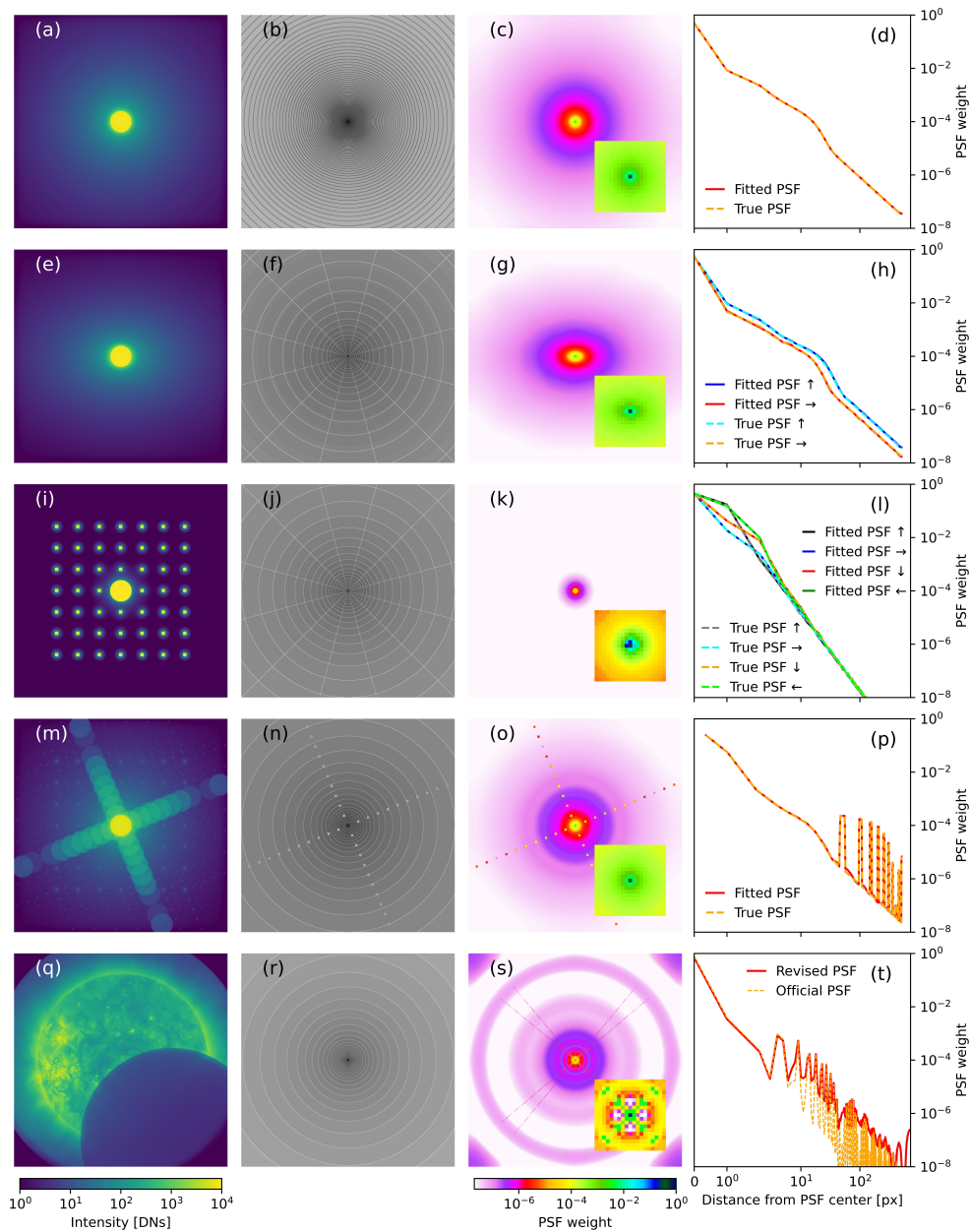}
    \caption{Five further use examples. The first column shows the observed partially occulted images, the second column the discretized PSFs where each segment corresponds to one PSF coefficient to fit, the third column the derived PSFs where the inset zooms into the PSF core, and the fourth column the derived and the true PSF weights along a slice through the center of the PSF in the horizontal direction for (d), in the horizontal and vertical directions for (h) and (l), and in the diagonal direction along the diffraction pattern for (p) and (t). (a)-(d)~We re-parametrize the PSF coefficients as a Gaussian function superimposed with a Lorentzian. (e)-(h)~We discretize the PSF into shell segments, which enables one to fit anisotropic PSFs such as ellipses. (i)-(l)~We discretize the PSF into shell segments to fit a PSF consisting of an Airy pattern with coma and astigmatism aberrations. (m)-(p)}~We give each location in the diffraction pattern its own coefficient, which enables one to fit a diffraction patterns. (q)-(t)~We revise the PSF of AIA, which observes the Sun.
    \label{fig:furtherpsfs}
\end{figure}

Here, we present 5~additional examples of how our algorithm can be used to determine the instrumental PSF. In the first example, we determine free parameters of a PSF whose functional form can be guessed. In the second example, we fit the weights of an elliptical PSF. In the third example, we derive the PSF for an Airy pattern with coma and astigmatism abberations. In the fourth example, we compute the weights in a PSF containing a diffraction pattern. And in the fifth example, we revise the PSF of a satellite imager. These examples are presented in Figure~\ref{fig:furtherpsfs}, whereby each row shows one example. The first column shows the observed illuminated occultation masks of these examples, the second column  shows the discretized PSFs, the third column shows the fitted PSFs, and the fourth column compares the weights of the fitted PSFs and the weights of the true PSFs along a PSF cross-section. 

In the first example, shown in Figure~\ref{fig:furtherpsfs}(a)-(d), we take as the true PSF the PSF of Sections~\ref{sec:showcase1} and~\ref{sec:showcase2}, defined by a Gaussian core and a Lorentzian tail. We select an occultation mask which contains in its center a single large hole with a radius of~$50$~pixels, and discretize the PSF into $200$~shells. We assume that the functional relationship of this PSF can already be guessed, e.g., from a previous fit of the PSF analog to Section~\ref{sec:showcase1}, and aim at determining the coefficients of the Gaussian core and Lorentzian tail. To do so, we parametrize the PSF coefficients $\text{psf}_S$ as a Gaussian function superposed with a Lorentzian, 
\begin{equation}
    \text{psf}_S(r) =  A\ \exp{ \left( -\frac{r^2}{2\ B^2} \right)} + \frac{C}{D + r^E}.
\end{equation}
When fitting the System of Equations \ref{eq:psf}, we fit for the coefficients $A$, $B$, $C$, $D$, and $E$. The fit results in 
\begin{align*}
    A &= \num{2.4282(5)e-4}, && B = \num{10.01(2)}, && C = \num{4.700(2)e-2}, \\ D &= 1.0802(4), && E = 1.99726085(5),
\end{align*}
where the one-sigma fitting uncertainties in the last digit are given in the parenthesis. These fitted values are very close to the true parameters $A=\num{2.3942e-4}$, $B= 10$, $C=\num{4.732e-2}  $, $D=1$, and $E=-2$. The MAPE of the fitted to the true PSF weights is \SI{0.4}{\percent}. 

In the second example, shown in  Figure~\ref{fig:furtherpsfs}(e)-(h), we define an elliptical true PSF by 
\begin{equation}
    \text{psf}(x, y) = \num{2.4e-4}\ \exp{\left(-\frac{0.25 x^2 +y^2}{200}\right)} + \frac{0.5}{1 + 61 (0.25 x^2+y^2)}, \label{eq:ellipticalPSF}
\end{equation}
where $x$ and $y$ are the horizontal and vertical distances from the PSF center. This PSF scatters \SI{54}{\percent} of the photons away from its center. The occultation mask contains a single large hole with a radius of~$50$~pixels. We discretize the PSF into shell segments, where we divide the $40$~cylindrical shells from Section~\ref{sec:showcase1} into segments having an angular width of \SI{30}{\degree}, resulting in $480$~PSF segments. 
We fit the PSF weights by randomly drawing \num{3000} lines from the system of equations and perform a multi-linear fit using the adaptive grid method described in Section~\ref{sec:implementation}. The result is presented in Figure~\ref{fig:furtherpsfs}(g), showing a smooth, elliptical PSF. In Figure~\ref{fig:furtherpsfs}(h), we plot the fitted and the true PSF weights along the horizontal and vertical axis through the PSF center. The weights of the fitted and true PSF agree very well. The MAPE between the fitted and true weights derived from all PSF segments is \SI{4.3}{\percent}.

In the third example, shown in Figure~\ref{fig:furtherpsfs}(i)-(l), we define the true PSF by an Airy pattern with coma and astigmatism aberrations. We create the true PSF by assuming a circular aperture, set the combined coma and astigmatism aberrations in the aperture function to a root-mean-square error of $0.1$~wavelength using Zernike polynomials, and compute numerically the true PSF as the Fourier transform of the aperture function. This PSF is asymmetric and scatters \SI{55}{\percent} of the photons away from its center. The occultation mask contains a large hole with a radius of~$50$~pixels and 48~pinholes each with a size of~$7$~pixel. We discretize the PSF into shell segments, where we divide the $40$~cylindrical shells from Section~\ref{sec:showcase1} into segments having an angular width of \SI{30}{\degree} each, resulting in $480$~PSF segments. We fit the PSF weights consecutively by performing multi-linear fits on subsystems of the system of equation as described in Section~\ref{sec:implementation}, and average the results of 10 fit repetitions. The resulting PSF  is presented in Figure~\ref{fig:furtherpsfs}(k). In Figure~\ref{fig:furtherpsfs}(l), we show the fitted and true PSF weights along axes starting from the PSF center towards the north, south, east, and west directions. The asymmetry of the true PSF due to the coma and astigmatism aberrations as well as the decreasing weights in the PSF tail are well represented. The MAPE of the fitted PSF from the true PSF is \SI{2.5}{\percent}.

In the fourth example, shown in Figure~\ref{fig:furtherpsfs}(m)-(p), we define the true PSF to have a Gaussian core, a Lorentzian tail, and a diffraction pattern, defined by
\begin{align}
    \text{psf}(x, y) &= \num{1.76e-4}\ \exp{\left(-\frac{x^2 +y^2}{200}\right)} + \frac{0.367}{1 + 61 (x^2+y^2)} + \nonumber \\
    &+ \begin{cases}
    \frac{0.367\ \sin ( 0.1\ \pi \ \sqrt{x^2 + y^2})}{0.1\ \pi \ \sqrt{x^2+y^2} } \quad &\text{for }  \tan^{-1}\frac{y}{x} = \pm \frac{\pi}{4} - \frac{\pi}{8} \\ & \quad \ \  \text{and} \ \left| \ \sqrt{x^2+y^2} \text{ mod } 18\text{~pixel}\ \right| \leq 4~\text{pixel}   \\
    0 & \text{else}
    \end{cases}.
\end{align}
This PSF scatters \SI{73}{\percent} of the photons away from its center. The occultation mask contains a large hole with a radius of~$50$~pixels and 48~pinholes each with a size of~$1$~pixel. We discretize the PSF into shells and additionally assign each location in the diffraction pattern its own segment. To fit for the PSF weights, we use a simple multi-linear fit. In the postprocessing step, only the shells are smoothed, and the fitted weights of the diffraction pattern are re-inserted afterwards.  The resulting PSF  is presented in Figure~\ref{fig:furtherpsfs}(o). In Figure~\ref{fig:furtherpsfs}(p), we show the fitted and true PSF weights in the direction through the diffraction pattern. The fitted PSF weights of the diffraction pattern, the Gaussian core, and the Lorentzian tail agree very well with the true PSF. The MAPE of the fitted PSF from the true PSF is \SI{1.4}{\percent}.

In the fifth example, shown in Figure~\ref{fig:furtherpsfs}(q)-(t), we revise the instrumental PSF for AIA, which observes the  solar atmosphere in $10$ filters at extreme ultraviolet, ultraviolet, and visible wavelength at a resolution of $4096\times 4096$ pixels and a cadence of \numrange{12}{24} seconds. The instrumental PSF is known to contain a diffraction pattern of two crosses originating from spectral filters, and the theoretical PSF weights in the diffraction pattern have been confirmed by studies of flares, i.e., strong localized energy outbreaks in the solar atmosphere which act as strong point sources. Nevertheless, in solar eclipse images, a small number of counts is still measured within the occulted area even after deconvolving the image with the instrumental PSF. Therefore, we assume that the weights in the PSF tail, responsible for long-distance scattered light, are underestimated. To revise the instrumental PSF, we use an eclipse image in the \SI{193}{\Angstrom} filter taken on 15 May 2012, where the eclipse serves as external occulter, and rebin the image to a resolution of $1024\times 1024$ pixels. We discretize the PSF into 40 shells, fit the missing PSF weights using Equation~\ref{eq:psf_revise}, and average the results of 10 fit repetitions. The result is presented in Figure~\ref{fig:furtherpsfs}(s), showing the original diffraction pattern superposed with the newly fitted smoothed shells. In Figure~\ref{fig:furtherpsfs}(t), we plot for the instrumental PSF and our revised PSF the shell-averaged weights of the PSF versus their distance from the PSF center. In both PSFs, the diffraction pattern is clearly visible as peaks in the PSF weights. However, in the revised PSF, the PSF weights in the PSF tail outside of the diffraction pattern are significantly larger. In the original image, the average intensity of the solar image was \SI{71}{DNs}, and the average counts in the eclipse region was \SI{1.3}{DNs}. Deconvolving the image with the PSF provided by the instrument team increases the average intensity of the solar image to \SI{72}{DNs} and reduces the counts in the eclipse region to~\SI{1.1}{DNs}. Deconvolving the image with our revised PSF increases the counts of the solar image to \SI{73}{DNs} and diminishes the counts in the eclipse region to \SI{0.18}{DNs}.

\section{Summary}  \label{sec:prospects}

We have presented a semi-empirical semi-blind algorithm that enables one to determine accurately the instrumental PSF of an imaging system from partially occulted images. Our algorithm converges towards the true PSF solution both in the core and the tail of the PSF, is easy to implement, noise resistant, and does not require a point source. Furthermore, the method enables one to fit an arbitrary PSF independent of the functional relationship of the PSF, and is only dependent on the initial segmentation chosen for the PSF.

Our algorithm combines both advantages of empirical algorithms and blind-deconvolution algorithms. Similar to blind-deconvolution algorithms, the utilization of entire images that are partially occulted enables one to derive the PSF in an automated manner. Using entire images further boosts the total number of photons available, which enables one to fit both the PSF core and the PSF tail simultaneously and accurately. Similar to empirical algorithms, we utilize the information on where the true image is zero. In our algorithm, this enables the algorithm to converge to the true solution.

We have tested our algorithm on six numerical use cases: a cylindrically symmetric PSF, a cylindrically symmetric PSF where the functional form of the PSF was provided in advance and only free parameters in the functional form had to be fit, an elliptical PSF, a PSF consisting of an Airy pattern with coma and astigmatism aberrations, a PSF containing a diffraction pattern, and the PSF of a real imager onboard a satellite. For these studies, we have used occultation masks that contain a large hole and multiple pinholes to provide the partial occultation for the first five cases and a solar eclipse as external occultation for the latter case. Typical runtimes of our algorithm for images containing $1024\times 1024$~pixels are \SI{15}{seconds} to \SI{4}{minutes} per iteration, depending on the numbers of parameters to fit, with typically less than five iterations required for convergence. The MAPE of the fitted to the true PSF coefficients was less than \SI{5}{\percent} for all test cases. 

Our algorithm works with any type of partially occulted or partially illuminated images. In addition to calibrating instruments in laboratory settings, another strength of our algorithm is in calibrating instruments for which calibration exposures are taken on a regular basis (e.g., for medical X-ray imagers), or for imagers which only need a one-time calibration and which take partially occulted images by chance. These latter include imagers on satellites, where the field of view might be partially blocked by debris or other satellites when looking down to Earth, or by solar eclipses when looking towards the Sun. One can also use images of stars and distant galaxies. Although their field of view is not externally occulted, the intensity is a priori known to be negligible for every pixel not containing a star. This makes our algorithm versatile for accurately determining the PSF of imagers in many diverse situations.

\begin{backmatter}
\bmsection{Funding}
This work has been supported, in part, by the NASA Heliophysics Living with a Star program through grant number 80NSSC20K0183 and by the German Research Foundation (DFG) through grant number 448336908.

\bmsection{Acknowledgments}
The solar image in Figure \ref{fig:furtherpsfs} is provided by courtesy of NASA/SDO and the AIA team.

\bmsection{Disclosures}
The authors declare no conflicts of interest.

\bmsection{Data availability}
A Python implementation of this algorithm and a library that creates the test cases presented here are available at https://github.com/stefanhofmeister/Deriving-PSFs-from-partially-occulted-images
\end{backmatter}

\bibliography{How to derive and revise point spread functions for imagers}

\begin{thebibliography}{10}
\newcommand{\enquote}[1]{``#1''}

\bibitem{krist1993}
J.~{Krist}, \enquote{{Tiny Tim : an HST PSF Simulator},} in \emph{Astronomical
  Data Analysis Software and Systems II,}  vol.~52 of \emph{Astronomical
  Society of the Pacific Conference Series} R.~J. {Hanisch}, R.~J.~V.
  {Brissenden}, and J.~{Barnes}, eds. (1993), p. 536.

\bibitem{krist1995}
J.~{Krist}, \enquote{{Simulation of HST PSFs using Tiny Tim},} in
  \emph{Astronomical Data Analysis Software and Systems IV,}  vol.~77 of
  \emph{Astronomical Society of the Pacific Conference Series} R.~A. {Shaw},
  H.~E. {Payne}, and J.~J.~E. {Hayes}, eds. (1995), p. 349.

\bibitem{hasan1995}
H.~{Hasan} and C.~J. {Burrows}, \enquote{{Telescope Image Modelling (TIM)},}
  {\protect\JournalTitle{Publications of the Astronomical Society of the
  Pacific}} \textbf{107}, 289 (1995).

\bibitem{jerius2000}
D.~{Jerius}, R.~H. {Donnelly}, M.~S. {Tibbetts}, R.~J. {Edgar}, T.~J. {Gaetz},
  D.~A. {Schwartz}, L.~P. {Van Speybroeck}, and P.~{Zhao}, \enquote{{Orbital
  measurement and verification of the Chandra X-ray Observatory's PSF},} in
  \emph{X-Ray Optics, Instruments, and Missions III,}  vol. 4012 of
  \emph{Society of Photo-Optical Instrumentation Engineers (SPIE) Conference
  Series} J.~E. {Truemper} and B.~{Aschenbach}, eds. (2000), pp. 17--27.

\bibitem{karovska2001}
M.~{Karovska}, S.~J. {Beikman}, M.~S. {Elvis}, J.~M. {Flanagan}, T.~{Gaetz},
  K.~J. {Glotfelty}, D.~{Jerius}, J.~C. {McDowell}, and A.~H. {Rots},
  \enquote{{The Chandra X-ray Observatory PSF Library},} in \emph{Astronomical
  Data Analysis Software and Systems X,}  vol. 238 of \emph{Astronomical
  Society of the Pacific Conference Series} J.~{Harnden}, F.~R., F.~A.
  {Primini}, and H.~E. {Payne}, eds. (2001), p. 435.

\bibitem{carter2003}
C.~{Carter}, M.~{Karovska}, D.~{Jerius}, K.~{Glotfelty}, and S.~{Beikman},
  \enquote{{ChaRT: The Chandra Ray Tracer},} in \emph{Astronomical Data
  Analysis Software and Systems XII,}  vol. 295 of \emph{Astronomical Society
  of the Pacific Conference Series} H.~E. {Payne}, R.~I. {Jedrzejewski}, and
  R.~N. {Hook}, eds. (2003), p. 477.

\bibitem{grigis2012}
P.~Grigis, Y.~Su, and M.~Weber, \enquote{{AIA PSF Characterization and Image
  Deconvolution},} Tech. rep., NASA, LMSAL, SAO (2012).

\bibitem{westergaard2012}
N.~J. {Westergaard}, K.~K. {Madsen}, N.~F. {Brejnholt}, J.~E. {Koglin}, F.~E.
  {Christensen}, M.~J. {Pivovaroff}, and J.~K. {Vogel}, \enquote{{NuSTAR
  on-ground calibration: I. Imaging quality},} in \emph{Space Telescopes and
  Instrumentation 2012: Ultraviolet to Gamma Ray,}  vol. 8443 of \emph{Society
  of Photo-Optical Instrumentation Engineers (SPIE) Conference Series}
  T.~{Takahashi}, S.~S. {Murray}, and J.-W.~A. {den Herder}, eds. (2012), p.
  84431X.

\bibitem{madsen2015}
K.~K. {Madsen}, F.~A. {Harrison}, C.~B. {Markwardt}, H.~{An}, B.~W.
  {Grefenstette}, M.~{Bachetti}, H.~{Miyasaka}, T.~{Kitaguchi}, V.~{Bhalerao},
  S.~{Boggs}, F.~E. {Christensen}, W.~W. {Craig}, K.~{Forster}, F.~{Fuerst},
  C.~J. {Hailey}, M.~{Perri}, S.~{Puccetti}, V.~{Rana}, D.~{Stern}, D.~J.
  {Walton}, N.~{J{\o}rgen Westergaard}, and W.~W. {Zhang},
  \enquote{{Calibration of the NuSTAR High-energy Focusing X-ray Telescope.}}
  {\protect\JournalTitle{The Astrophysical Journal Supplement Series}}
  \textbf{220}, 8 (2015).

\bibitem{hiraoka1990}
Y.~{Hiraoka}, J.~W. {Sedat}, and D.~A. {Agard}, \enquote{{Determination of
  three-dimensional imaging properties of a light microscoope system. Partial
  confocal behavior in epifluorescence microscopy.}}
  {\protect\JournalTitle{Biophysical journal}} \textbf{57,2}, 325--33 (1990).

\bibitem{shaw1991}
P.~J. Shaw and D.~J. Rawlins, \enquote{The point-spread function of a confocal
  microscope: its measurement and use in deconvolution of 3-d data,}
  {\protect\JournalTitle{Journal of Microscopy}} \textbf{163}, 151--165 (1991).

\bibitem{boutet2001}
J.~{Boutet de Monvel}, S.~{Le Calvez}, and M.~Ulfendahl, \enquote{Image
  restoration for confocal microscopy: Improving the limits of deconvolution,
  with application to the visualization of the mammalian hearing organ,}
  {\protect\JournalTitle{Biophysical Journal}} \textbf{80}, 2455--2470 (2001).

\bibitem{Juskaitis2006}
R.~Ju{\v{s}}kaitis, \emph{Measuring the Real Point Spread Function of High
  Numerical Aperture Microscope Objective Lenses} (Springer US, Boston, MA,
  2006), pp. 239--250.

\bibitem{Jizhou2018}
J.~Li, F.~Xue, F.~Qu, Y.-P. Ho, and T.~Blu, \enquote{On-the-fly estimation of a
  microscopy point spread function,} {\protect\JournalTitle{Opt. Express}}
  \textbf{26}, 26120--26133 (2018).

\bibitem{tikhonov1977}
A.~N. Tikhonov and V.~Y. Arsenin, \emph{Solutions of ill-posed problems} (V. H.
  Winston \& Sons, Washington, D.C.: John Wiley \& Sons, New York, 1977).
  Translated from the Russian, Preface by translation editor Fritz John,
  Scripta Series in Mathematics.

\bibitem{rudin1992}
L.~I. Rudin, S.~Osher, and E.~Fatemi, \enquote{Nonlinear total variation based
  noise removal algorithms,} {\protect\JournalTitle{Physica D: Nonlinear
  Phenomena}} \textbf{60}, 259--268 (1992).

\bibitem{rudin1994}
L.~Rudin and S.~Osher, \enquote{Total variation based image restoration with
  free local constraints,} in \emph{Proceedings of 1st International Conference
  on Image Processing,}  vol.~1 (1994), pp. 31--35 vol.1.

\bibitem{nakashima2003}
P.~Nakashima and A.~Johnson, \enquote{Measuring the psf from aperture images of
  arbitrary shape—an algorithm,} {\protect\JournalTitle{Ultramicroscopy}}
  \textbf{94}, 135--148 (2003).

\bibitem{levenberg1944}
K.~LEVENBERG, \enquote{A method for the solution of certain non-linear problems
  in least squares,} {\protect\JournalTitle{Quarterly of Applied Mathematics}}
  \textbf{2}, 164--168 (1944).

\bibitem{marquardt1963}
D.~W. Marquardt, \enquote{An algorithm for least-squares estimation of
  nonlinear parameters,} {\protect\JournalTitle{Journal of the Society for
  Industrial and Applied Mathematics}} \textbf{11}, 431--441 (1963).

\bibitem{Tikhonov1943}
A.~N. Tikhonov, \enquote{On the stability of inverse problems,}
  {\protect\JournalTitle{Proceedings of the USSR Academy of Sciences}}
  \textbf{39}, 195--198 (1943).

\bibitem{Tibshirani1996}
R.~Tibshirani, \enquote{Regression shrinkage and selection via the lasso,}
  {\protect\JournalTitle{Journal of the Royal Statistical Society. Series B
  (Methodological)}} \textbf{58}, 267--288 (1996).

\bibitem{Lee2011}
J.-H. Lee and Y.-S. Ho, \enquote{High-quality non-blind image deconvolution
  with adaptive regularization,} {\protect\JournalTitle{J. Vis. Commun. Image
  Represent.}} \textbf{22}, 653--663 (2011).

\bibitem{lemen2012}
J.~R. {Lemen}, A.~M. {Title}, D.~J. {Akin}, P.~F. {Boerner}, C.~{Chou}, J.~F.
  {Drake}, D.~W. {Duncan}, C.~G. {Edwards}, F.~M. {Friedlaender}, G.~F.
  {Heyman}, N.~E. {Hurlburt}, N.~L. {Katz}, G.~D. {Kushner}, M.~{Levay}, R.~W.
  {Lindgren}, D.~P. {Mathur}, E.~L. {McFeaters}, S.~{Mitchell}, R.~A. {Rehse},
  C.~J. {Schrijver}, L.~A. {Springer}, R.~A. {Stern}, T.~D. {Tarbell}, J.-P.
  {Wuelser}, C.~J. {Wolfson}, C.~{Yanari}, J.~A. {Bookbinder}, P.~N.
  {Cheimets}, D.~{Caldwell}, E.~E. {Deluca}, R.~{Gates}, L.~{Golub}, S.~{Park},
  W.~A. {Podgorski}, R.~I. {Bush}, P.~H. {Scherrer}, M.~A. {Gummin},
  P.~{Smith}, G.~{Auker}, P.~{Jerram}, P.~{Pool}, R.~{Soufli}, D.~L. {Windt},
  S.~{Beardsley}, M.~{Clapp}, J.~{Lang}, and N.~{Waltham}, \enquote{{The
  Atmospheric Imaging Assembly (AIA) on the Solar Dynamics Observatory (SDO)},}
  {\protect\JournalTitle{Solar Physics}} \textbf{275}, 17--40 (2012).

\bibitem{pesnell2012}
W.~D. {Pesnell}, B.~J. {Thompson}, and P.~C. {Chamberlin}, \enquote{{The Solar
  Dynamics Observatory (SDO)},} {\protect\JournalTitle{Solar Physics}}
  \textbf{275}, 3--15 (2012).

\end{thebibliography}

\clearpage

\end{document}